\documentclass[a4paper]{article}
\usepackage{latexsym, epsfig, subfigure}
\begin{document}
\title{Bayesian analysis of finite population sampling in multivariate co-exchangeable structures with separable covariance matrices}
\author{}  
\date{}
\maketitle
\vspace{-10mm}
\begin{center} BY SIMON C. SHAW \\
\textit{Department of Mathematical Sciences, University of Bath, Bath, BA2 7AY, U.K.} \\ s.shaw@bath.ac.uk \\
AND MICHAEL GOLDSTEIN \\
                          \textit{Department of Mathematical Sciences, Durham University, Science Laboratories, \\
                          South Road, Durham, DH1 3LE, U.K.} \\
michael.goldstein@durham.ac.uk \end{center}
\newenvironment{summary}{\begin{quote} \begin{center}
\large{\textbf{ABSTRACT}}  \end{center}}{\end{quote}}
\newtheorem{Theo}{Theorem}
\newtheorem{Lem}{Lemma}
\newtheorem{Cor}{Corollary}   
\newtheorem{Def}{Definition}
\newtheorem{Assum}{Assumption}
\newcommand{\cip}{\mbox{$\perp\!\!\!\perp$}}
\newcommand{\spec}[1]{\ensuremath{\lfloor #1 \rceil}}
\newcommand{\lincom}[1]{\ensuremath{\langle #1 \rangle}}

\begin{summary}
We explore the effect of finite population sampling in design problems with many variables cross-classified in many ways. In particular, we investigate designs where we wish to sample individuals belonging to different groups for which the underlying covariance matrices are separable between groups and variables. We exploit the generalised conditional independence structure of the model to show how the analysis of the full model can be reduced to an interpretable series of lower dimensional problems. The types of information we gain by sampling are identified with the orthogonal canonical directions. We first solve a variable problem, which utilises the powerful properties of the adjustment of second-order exchangeable vectors, which has the same qualitative features, represented by the underlying canonical variable directions, irrespective of chosen group, population size or sample size. We then solve a series of group problems which in a balanced design reduce to the sampling of second-order exchangeable vectors. If the population sizes are finite then the qualitative and quantitative features of each group problem will depend upon the sampling fractions in each group, mimicking the infinite problem when the sampling fractions in each group are the same. 
\end{summary}
\textit{Some key words}: Bayes linear methods; canonical directions; canonical resolutions; second-order exchangeability; finite population sampling; separable covariance, dimension reduction. 
\section{Introduction \label{sec:Intro}}

Increasingly, statisticians are confronted by large multidimensional problems. The modeling and analysis of such problems is challenging: we seek to model as accurately as possible, within our limited specification capabilities, whilst still trying to maintain a tractable analysis. In this paper, we investigate designs where we wish to sample individuals belonging to different groups, exploiting the properties of sampling second-order exchangeable vectors. In particular, we consider the effect of judging that the individual covariance matrices are separable between groups and variables so that they can be represented as the product of a covariance matrix over groups and a covariance matrix over variables. 

Models utilising separable covariance matrices have been used in a number of applications. For example, in multivariate repeated measures data utilising a covariance matrix for individuals that is separable between measured characteristics and time points (Naik and Rao, 2001\nocite{Naik/Rao:2001}; Chaganty and Naik, 2002\nocite{Chaganty/Naik:2002}; Roy and Khattree, 2005\nocite{Roy/Khattree:2005}; Mitchell et al, 2006\nocite{Mitchell/etal:2006}); in spatio-temporal models separating the space and time effects (Mardia and Goodall, 1993\nocite{Mardia/Goodall:1993}; Huizenga et al, 2002\nocite{Huizenga/etal:2002}; Fuentes, 2006\nocite{Fuentes:2006}; Genton, 2007\nocite{Genton:2007}); in the modelling of emulators for computer experiments using a separation between inputs and outputs (Sacks et al, 1989\nocite{Sacks/etal:1989}; Currin et al, 1991\nocite{Currin/etal:1991}; Rougier, 2008\nocite{Rougier:2008}). Wang and West (2009)\nocite{Wang/West:2009} develop a Bayesian analysis of matrix normal graphical models; if a $n \times p$ matrix $X$ follows a matrix normal distribution then the vectorisation of $X$, $vec(X)$, follows a multivariate normal distribution whose covariance matrix is separable into a $n \times n$ covariance matrix and a $p \times p$ covariance matrix. As we note in Section \ref{sec:3.1.1}, the prior for the model we utilise could be viewed as a multivariate normal distribution of this type. Hoff (2011)\nocite{Hoff:2011} looks at extensions to the matrix normal model to incorporate multidimensional data arrays.

We consider a model where individuals in each group belong to second-order exchangeable populations and that they are co-exchangeable between groups with the individual covariance matrices being judged to be separable between groups and variables. Given a sample of individuals from each group updating is performed using Bayes linear methods. As an illustrative example of the theory we develop we consider the performance of a collection of individuals on an examination paper which contains a number of compulsory questions with each paper being marked by one of a series of markers. The examiner has a number of questions of interest such as whether the questions were at the appropriate level of difficulty and whether each marker was marking to the same standard. Individuals with the same marker are judged to be second-order exchangeable and co-exchangeable between markers. It is judged that the individual covariance matrices are separable between markers and questions.

Shaw and Goldstein (1999)\nocite{Shaw/Goldstein:1999} considered a related model where the population in each group was judged to be infinite and separable covariance matrices assumed over underlying mean and residual vectors. They showed that the full problem may be decomposed into smaller subspaces over which the update has the same qualitative features: the full solution may be obtained by analysing an interpretable variable problem, representing sampling from an infinite second-order exchangeable population, and an interpretable group problem.

However, the judgement of infinite populations is often a simplifying one to the finite reality. It is important to understand when a full accounting of the finiteness is significant, for example whether or not the sampling fraction is ignorable. Shaw and Goldstein (2012)\nocite{Shaw/Goldstein:2012} consider multivariate Bayesian sampling from a finite population. Developing the work of Goldstein and Wooff (1998)\nocite{Goldstein/Wooff:1998} they show how, for the sampling of second-order exchangeable vectors, the familiar univariate finite population corrections naturally generalise to individual quantities in the multivariate population. The canonical directions share the same coordinate representation for all sample sizes and, for equally defined individuals, all population sizes. 

The paper proceeds as follows. In Section \ref{sec:BLB} we briefly review updating using Bayes linear methods and how to utilise the resolution transform to perform a canonical analysis. In Section \ref{sec:two} we introduce the model which represents multivariate co-exchangeable structures, or groups, with separable covariance matrices.  We explore the model first in the case where we sample from infinite populations, Section \ref{sec:sampinf}, and then in Section \ref{sec:finite} when populations are judged to be finite. We demonstrate that in both cases the full problem may be tackled by reducing it into smaller interpretable subspaces. Firstly, a variable problem which represents sampling from a second-order exchangeable population where the finite problem shares the same qualitative features as the infinite problem. Secondly, a series of group problems which in the special case of a balanced design reduce to sampling from an exchangeable second-order population. In the finite setting, both the qualitative and quantitative features of the group problem differ from the infinite case although when the sampling fraction is the same in each group, the qualitative features of the finite problem match the infinite with simple population corrections applied to account for the quantitative differences. In particular, in the case of the balanced design with equal population sizes, the group problems reduce to one of finite second-order exchangeable population sampling. The theory is illustrated by a simple example concerning examination data.

\section{Bayes linear methods and canonical structure \label{sec:BLB}}
For a collection of random quantities $\mathcal{B} = \{B_{1}, \ldots, B_{r}\}$ we denote by $\langle \mathcal{B} \rangle$ the collection of linear combinations, $Y = \sum_{i=1}^{r} h_{i} B_{i}$. For notational convenience, we also consider $\mathcal{B}$ as the $r \times 1$ vector with $Y = h^{T} \mathcal{B}$. Suppose that we observe the values of a further collection of random quantities $\mathcal{D} = \{D_{1}, \ldots, D_{s}\}$ and that we have specified prior means, variances and covariances for the collection $\mathcal{B} \cup \mathcal{D}$. We intend to revise our beliefs about $\mathcal{B}$ given the values of $\mathcal{D}$. For any $Y, Z \in \langle \mathcal{B} \rangle$ we utilise the Bayes linear methodology to obtain $E_{\mathcal{D}}(Y)$, the adjusted expectation of $Y$ given $\mathcal{D}$, $Var_{\mathcal{D}}(Y)$ the adjusted variance of $Y$ given $\mathcal{D}$, and $Cov_{\mathcal{D}}(Y, Z)$, the adjusted covariance between $Y$ and $Z$ given $\mathcal{D}$. Similarly for the collection $\mathcal{B}$ we can obtain the vector $E_{\mathcal{D}}(\mathcal{B})$, the adjusted expectation, and matrix $Var_{\mathcal{D}}(\mathcal{B})$, the adjusted variance. They are calculated as
\begin{eqnarray}
E_{\mathcal{D}}(\mathcal{B}) & = & E(\mathcal{B}) + Cov(\mathcal{B}, \mathcal{D})Var^{\dagger}(\mathcal{D})\{\mathcal{D} - E(\mathcal{D})\} \label{eq:adj1} \\
Var_{\mathcal{D}}(\mathcal{B}) & = & Var(\mathcal{B}) - Cov(\mathcal{B}, \mathcal{D})Var^{\dagger}(\mathcal{D})Cov(\mathcal{D},\mathcal{B}) \label{eq:adj2}
\end{eqnarray}
where $Var^{\dagger}(\mathcal{D})$ is the Moore-Penrose generalised inverse of $Var(\mathcal{D})$. From (\ref{eq:adj1}) and (\ref{eq:adj2}) we may obtain the adjusted second-order beliefs for every $Y, Z \in \langle \mathcal{B} \rangle$ and refer to such a case as the adjustment of $\langle \mathcal{B} \rangle$ given $\mathcal{D}$. Chapter 3 of Goldstein and Wooff (2007) \nocite{Goldstein/Wooff:2007} gives a detailed explanation of the adjustment of beliefs in the Bayes linear paradigm. Note that if $\mathcal{B}$ and $\mathcal{D}$ are jointly normally distributed then the adjusted quantities coincide with the usual definitions of conditional expectation; see Hartigan (1969; p447).
\nocite{Hartigan:1969}

We also want to gain insight into the types, and strength, of information given by the observation $\mathcal{D}$. This can be done by performing a canonical analysis; see Section 3.9 of Goldstein and Wooff (2007) \nocite{Goldstein/Wooff:2007} for the general approach and Goldstein (1981) \nocite{Goldstein:1981} for the fundamental geometric interpretation. For each $Y \in \langle \mathcal{B} \rangle$ the resolution $Res_{\mathcal{D}}(Y) = 1 - \{Var_{\mathcal{D}}(Y)/Var(Y)\}$ is a simple scale-free measurement of the effect of $\mathcal{D}$ on $Y$. 
\begin{Def} (Goldstein and Wooff, 2007)\nocite{Goldstein/Wooff:2007} The $j$th canonical direction for the adjustment of $\mathcal{B}$ by $\mathcal{D}$ is the linear combination $Y_{j}$ which maximises $Res_{\mathcal{D}}(Y)$ over all elements $Y \in \langle \mathcal{B} \rangle$ with non-zero prior variance which are uncorrelated a priori with $Y_{1}, \ldots, Y_{j-1}$. The value $\lambda_{j} = Res_{\mathcal{D}}(Y_{j})$ is termed the $j$th canonical resolution.
\end{Def}
The $Y_{j}$ are typically scaled to have prior variance one and can be centred to have prior expectation zero. The number of canonical directions is equal to the rank of the variance matrix of the elements of $\mathcal{B}$. 
\begin{Theo} (Goldstein and Wooff, 2007)\nocite{Goldstein/Wooff:2007} The $j$th canonical resolution for the adjustment of $\mathcal{B}$ by $\mathcal{D}$ is the $j$th largest eigenvalue, $\lambda_{j}$, of the resolution transform matrix
\begin{eqnarray}
T_{\mathcal{B} : \mathcal{D}} & = & Var^{\dagger}(\mathcal{B})Cov(\mathcal{B}, \mathcal{D})Var^{\dagger}(\mathcal{D})Cov(\mathcal{D},\mathcal{B}). \label{eq:adj3}
\end{eqnarray}
The $j$th canonical direction is the linear combination $h_{j}^{T}\mathcal{B}$, where $h_{j}$ is the eigenvector corresponding to $\lambda_{j}$.
\end{Theo}
For simplicity of exposition, in this paper we shall assume that all variance matrices are of full rank. As in Goldstein and Wooff (1998),\nocite{Goldstein/Wooff:1998} if we do not have invertibility, we obtain corresponding results over the linear span of the columns of the corresponding matrices. Notice that $1 \geq \lambda_{1} \geq \cdots \geq \lambda_{r} \geq 0$ and that each $Y$ may be expressed as $Y = \sum_{j=1}^{r} Cov(Y, Y_{j})Var^{-1}(Y_{j}) Y_{j}$. The resolution transform summarises all the effects of the adjustment over $\langle \mathcal{B} \rangle$ given $\mathcal{D}$ as for any $Y \in \langle \mathcal{B} \rangle$ we have $Var_{\mathcal{D}}(Y) = \sum_{j=1}^{r} Cov^{2}(Y, Y_{j})Var^{-1}(Y_{j})(1 - \lambda_{j})$ leading to the resolution partition
\begin{eqnarray}
Res_{\mathcal{D}}(Y) & = & \sum_{j=1}^{r} Corr^{2}(Y, Y_{j})\lambda_{j} \label{eq:respart}
\end{eqnarray}
 where $\sum_{j=1}^{r} Corr^{2}(Y, Y_{j}) = 1$. The resolution partition shows that each $Y$ can be expressed as a sum of the $Y_{j}$ about each of which $\mathcal{D}$ provides progressively less information: we expect to learn most about elements of $\langle \mathcal{B} \rangle$ which have strong correlations with the early $Y_{j}$.

\section{Multivariate co-exchangeable structures with separable covariance matrices \label{sec:two}}
\subsection{The model  \label{sec:model}}

We consider sampling individuals who can be identified as belonging to one of $g_{0}$ groups. For each individual we wish to make the same series of measurements $\mathcal{C} = \{X_{1}, \ldots, X_{v_{0}}\}$ and let $\mathcal{C}_{gi} = \{X_{g1i}, \ldots, X_{gv_{0}i}\}$ denote the measurements for the $i$th individual in the $g$th group. 
\begin{Assum} \label{assum1}
We judge that individuals in each group are second-order exchangeable and that they are co-exchangeable (Goldstein, 1986)\nocite{Goldstein:1986} across groups. For all $g \neq h$, $i \neq j$, $k$ our second-order specifications thus take the form
\begin{eqnarray*}
E(\mathcal{C}_{gi}) \ = \ \mu_{g}; \ Var(\mathcal{C}_{gi}) \ = \ D_{g}; \ Cov(\mathcal{C}_{gi}, \mathcal{C}_{gj}) \ = \ C_{gg}; \ Cov(\mathcal{C}_{gi}, \mathcal{C}_{hk}) \ = \ C_{gh}.
\end{eqnarray*}
\end{Assum}
Note that Assumption \ref{assum1}, without placing too harsh a constraint upon our beliefs, produces a prior specification of the model through a small number of specifications over observable quantities. For beliefs within the same group, all that is required is the consideration of the relationship between two individuals and for beliefs between two groups the relationship between a single individual from each group. In particular, the number of required specifications does not depend upon the population sizes within each group. Assumption \ref{assum1} provides the most general case of a model formed from second-order exchangeability and co-exchangeability. When considering extensions from exchangeability to partial exchangeability, de Finetti (1959)\nocite{deFinetti:1959} noted that ``passing directly to the most general case would be to renounce all possibility of illuminating the varied aspects of the question that merit interest.'' We follow his guidance in this second-order setting by considering a further simplification.
\begin{Assum} \label{assum2}
We judge that the individual covariance matrices are separable between groups and variables. For all $g$, $h$, $i$, $j$ we have
\begin{eqnarray*}
Cov(\mathcal{C}_{gi}, \mathcal{C}_{hj}) & = & \left\{\begin{array}{ll}
\gamma_{g}D & g = h, i = j; \\
\alpha_{gh}C & \mbox{otherwise}
\end{array}\right. 
\end{eqnarray*}
where $D = (d_{vw})$, $C = (c_{vw})$ are general $v_{0} \times v_{0}$ positive definite matrices, the $\alpha_{gh}$ are such that the $g_{0} \times g_{0}$ matrix $A = (\alpha_{gh})$ is positive definite and, for all $g$, $\gamma_{g} > 0$. We additionally form the $g_{0} \times g_{0}$ matrices $\hat{A} = diag(\alpha_{11}, \ldots, \alpha_{g_{0}g_{0}})$ and $B = diag(\gamma_{1}, \ldots, \gamma_{g_{0}})$.
\end{Assum}
We initially assume that the number of individuals in each group is (potentially) infinite. Then, see Goldstein(1986)\nocite{Goldstein:1986}, a consequence of the judgement of second-order exchangeability in Assumption \ref{assum1} is the representation theorem: we may write
\begin{eqnarray}
\mathcal{C}_{gi} & = & \mathcal{M}(\mathcal{C}_{g}) + \mathcal{R}_{i}(\mathcal{C}_{g}) \label{eq:as1}
\end{eqnarray}
 where $\mathcal{M}(\mathcal{C}_{g})$ is the limit, in mean square, of $\frac{1}{n_{g}} \sum_{i=1}^{n_{g}} \mathcal{C}_{gi}$ as $n_{g} \rightarrow \infty$. $\mathcal{M}(\mathcal{C}_{g}) = \{\mathcal{M}(X_{g1}), \ldots$, $\mathcal{M}(X_{gv_{0}})\}$ is the underlying $g$th group population mean vector and $\mathcal{R}_{i}(\mathcal{C}_{g}) = \mathcal{C}_{gi} -  \mathcal{M}(\mathcal{C}_{g})$ the residual vector for the $i$th individual in the $g$th group. For all $g$, $h$, $i$, $\mathcal{R}_{i}(\mathcal{C}_{g})$ is uncorrelated with $\mathcal{M}(\mathcal{C}_{h})$ whilst all of the residual vectors are mutually uncorrelated. From the specifications given by Assumptions \ref{assum1} and \ref{assum2} we have $E(\mathcal{M}(\mathcal{C}_{g})) = \mu_{g}$, $Var(\mathcal{M}(\mathcal{C}_{g})) = \alpha_{gg}C$, $Cov(\mathcal{M}(\mathcal{C}_{g}), \mathcal{M}(\mathcal{C}_{h})) = \alpha_{gh}C$ and $Var(\mathcal{R}_{i}(\mathcal{C}_{g})) = \gamma_{g}D - \alpha_{gg}C$. We collect the group population mean vectors together as $\mathcal{M}(\mathcal{C}) = \{\mathcal{M}(\mathcal{C}_{1}), \ldots, \mathcal{M}(\mathcal{C}_{g_{0}})\}$ so that $Var(\mathcal{M}(\mathcal{C})) =  A \otimes C$, the direct product of $A$ and $C$; see Searle et al.\ (1992)\nocite{Searle/etal:1992} for further details and properties of the direct product.
\subsubsection{Relationship with multivariate normal modelling \label{sec:3.1.1}}
The results that we develop for the model defined by Assumptions \ref{assum1} and \ref{assum2} coincide with those for the model when the joint distribution of any collection of individuals is multivariate normal with the same second-order structure. In particular, for each $g$ and any possible sequence length $m_{g}$, $\mathcal{C}_{g1}, \ldots, \mathcal{C}_{gm_{g}}$ are exchangeable with joint distribution $N(\eta_{g}, \Sigma_{g})$ where $\eta_{g} = 1_{m_{g}} \otimes \mu_{g}$, with $1_{m_{g}}$ denoting the $m_{g} \times 1$ vector of ones, and $\Sigma_{g}$ is the $m_{g}v_{0} \times m_{g}v_{0}$ block matrix with $(i,i)$th block $\alpha_{g}D$ and $(i, j)$th block $\alpha_{gg}C$, $i \neq j = 1, \ldots, m_{g}$. There is a model of partial exchangeability between groups. For a sequence length of $m_{g}$ in the $g$th group, $g = 1, \ldots, g_{0}$, the joint distribution of the individuals is $N(\eta, \Sigma)$ where $\eta = [\eta_{1}^{T} \ldots \eta_{g_{0}}^{T}]^{T}$ and $\Sigma$ is the $(\sum_{g=1}^{g_{0}} m_{g})v_{0} \times (\sum_{g=1}^{g_{0}} m_{g})v_{0}$ block matrix with $(g,g)$th block $\Sigma_{g}$ and $(g, h)$th block $1_{m_{g}}1_{m_{h}}^{T} \otimes \alpha_{gh}C$, $g \neq h = 1, \ldots, g_{0}$.

If the number of individuals in each group is infinite then, via de Finetti's representation theorem, we may introduce the $g_{0}v_{0}$ parameters $\mathcal{M}(\mathcal{C})$ where, conditional upon $\mathcal{M}(\mathcal{C})$, for each $g$, the $\mathcal{C}_{gi}$ are independent and identically distributed $N(\mathcal{M}(\mathcal{C}_{g}), \gamma_{g}D-\alpha_{gg}C)$, and for $g \neq h$ $\mathcal{C}_{gi}$ and $\mathcal{C}_{hj}$ are independent. The prior $\mathcal{M}(\mathcal{C}) \sim N(\mu, A \otimes C)$ where $\mu = [\mu_{1}^{T} \ldots \mu_{g_{0}}^{T}]^{T}$. Equivalently, following the notation of Wang and West (2009)\nocite{Wang/West:2009}, if we form the $v_{0} \times g_{0}$ matrix $\widehat{\mathcal{M}}(\mathcal{C})$ whose $(v, g)$th entry is $\mu(X_{gv})$, then $\widehat{\mathcal{M}}(\mathcal{C})$ follows a matrix normal distribution with mean matrix $M$, where $vec(M) = \mu$, column variance matrix $C$ and row variance matrix $A$. 

\subsubsection{Alternative model formulation\label{sec:3.1.2}}
When considering a simplification to the model obtained by Assumption \ref{assum1}, instead of Assumption \ref{assum2} Shaw and Goldstein (1999)\nocite{Shaw/Goldstein:1999} impose separability between groups and variables by directly specifying $Var(\mathcal{M}(\mathcal{C})) = A \otimes C$ and $Var(\mathcal{R}_{i}(\mathcal{C})) = \beta_{g}E$ for all $g$, $i$ where $E$ is a general $v_{0} \times v_{0}$ positive definite matrix and $\beta_{g} > 0$. The corresponding beliefs over individuals are then deduced using the representation theorem, (\ref{eq:as1}). Note that this approach involves a specification over unobservable random quantities whose existence assumes an infinite population in each group. However, it is much more natural for a subjective Bayes analysis to impose separability on judgements about observable individuals and deduce the corresponding properties for the mean and residual quantities. This is the approach, via Assumption \ref{assum2}, in this paper and is even more relevant when, unlike Shaw and Goldstein (1999)\nocite{Shaw/Goldstein:1999}, in Section \ref{sec:finite} we wish to consider the population sizes as being finite so that the representation given by (\ref{eq:as1}) does not hold. In certain special cases, for example if $\alpha_{gg}/\gamma_{g}$ does not depend upon $g$, the two models coincide. Thus, the results for the model of Shaw and Goldstein (1999)\nocite{Shaw/Goldstein:1999} can be obtained as a special case of the model in this paper as we shall highlight in Corollary \ref{cor:prop} and the final paragraphs of Sections \ref{sec:groups} and \ref{sec:fingroups}.

\subsection{Examination data example\label{sec:ex1}}

To illustrate our theoretical development we consider an example concerning a first year university examination paper in probability and statistics which comprises of two sections. Section A consists of five short compulsory questions, each marked out of six, meant to test a student's engagement with the course whilst Section B has three long compulsory questions, each marked out of 15, which are designed to stretch the student. Under university regulations, as the examination does not count towards the student's final degree classification, it may be marked by graduate students rather than the examiner. In this case, three such students are available. The examiner has a number of questions of interest which include whether or not the questions in each section were of approximately similar difficulty and whether or not the markers marked to the same standard. 

The examiner denotes by $X_{gvi}$ the mark on the $v$th question by the $i$th candidate marked by the $g$th marker. Thus, $\{X_{g1i}, \ldots, X_{g5i}\}$ and $\{X_{g6i}, \ldots, X_{g8i}\}$ are, respectively, the corresponding Section A and Section B marks. Each script is anonymous and the examiner judges that individuals are second-order exchangeable within markers and co-exchangeable across markers so that Assumption \ref{assum1} holds. He also believes that, as the markers have all received similar training and briefings on the marking scheme, there should be little difference between them and any differences shouldn't depend upon the specific question so that Assumption \ref{assum2} holds. Consequently, for the (co)variance structure, his specifications reduce to two $8 \times 8$ positive definite matrices ($D$ and $C$), and two $3 \times 3$ positive definite matrices ($A$ and $B$) one of which is diagonal. 

In considering these specifications, the examiner judges that the questions are second-order exchangeable within sections and co-exchangeable across sections. He also believes that there should be little difference between the markers as they have all received similar training and briefings on the marking scheme and thus judges his beliefs to be second-order exchangeable over groups. These further assumptions impose additional structure upon the matrices $D$, $C$, $A$ and $B$ which, as we will see in Sections and \ref{sec:exvar} and \ref{sec:exinf}, will allow the explicit derivation of the canonical structure. Letting $I_{p}$ and $J_{p, q}$ respectively denote the $p \times p$ identity matrix and the $p \times q$ matrix of ones, the examiner specifies for all $g \neq h$, $i \neq j$, $k$
\begin{eqnarray*}
E(\mathcal{C}_{gi}) = [4J_{1,5} \ 7.5J_{1,3}]^{T}, \ Var(\mathcal{C}_{gi}) = D, \ Cov(\mathcal{C}_{gi}, \mathcal{C}_{gj}) = C, \ Cov(\mathcal{C}_{gi}, \mathcal{C}_{hk}) = 0.85C
\end{eqnarray*}
where
\begin{eqnarray*}
D = \left(\begin{array}{cc} I_{5}+2J_{5,5} & 2.75J_{5,3} \\
2.75J_{3, 5} & 4.5I_{3}+9.5J_{3, 3} \end{array} \right), \ C = \left(\begin{array}{cc} 0.8I_{5}+0.2J_{5,5} & 0.5J_{5,3} \\
0.5J_{3, 5} & 2.15I_{3}+0.85J_{3, 3} \end{array} \right) 
\end{eqnarray*}
so that $A = (0.15I_{3}+0.85J_{3, 3})$, $\hat{A} = I_{3}$, and $B = I_{3}$.

Initially, the examiner assumes that the individuals are drawn from (potentially) infinite populations representing, see (\ref{eq:as1}), in the univariate setting $X_{gvi} = \mathcal{M}(X_{gv}) + \mathcal{R}_{i}(X_{gv})$ or as vectors $\mathcal{C}_{gi} = \mathcal{M}(\mathcal{C}_{g}) + \mathcal{R}_{i}(\mathcal{C}_{g})$. He will be interested in learning about linear combinations, over both $g$ and $v$, of the $\mathcal{M}(X_{gv})$ given samples of the $X_{gvi}$. For example, $TotA_{g} = \sum_{v=1}^{5} \mathcal{M}(X_{gv})$, $TotB_{g} = \sum_{v=6}^{8} \mathcal{M}(X_{gv})$, $Tot_{g} = TotA_{g} + TotB_{g}$ denote, respectively, the underlying script total on Section A, on Section B, and overall for the $g$th marker. Similarly, $TotA_{1}-TotA_{2}$, $TotB_{1}-TotB_{2}$, $Tot_{1}-Tot_{2}$ denote, respectively, the difference in the underlying totals for Section A, Section B, and overall between the first and second marker.

\section{Sampling from infinite populations: adjustment of $\langle \mathcal{M}(\mathcal{C}) \rangle$ given $\mathcal{C}(N)$ \label{sec:sampinf}}
We wish to consider the effect of a sample of individuals on individual quantities in the population mean collection $\mathcal{M}(\mathcal{C})$. Suppose that we sample $n_{g}$ individuals in the $g$th group and let $\mathcal{C}_{g}(n_{g}) = \{\mathcal{C}_{g1}, \ldots, \mathcal{C}_{gn_{g}}\}$ denote the measurements of the individuals sampled in the $g$th group; for notational simplicity we have used the labelling convention that we sample the first $n_{g}$ individuals in the $g$th group. We collect the sample sizes together into the matrix $N = diag(n_{1}, \ldots, n_{g_{0}})$. The total collection of measurements are denoted by $\mathcal{C}(N) = \{\mathcal{C}_{1}(n_{1}), \ldots, \mathcal{C}_{g_{0}}(n_{g_{0}})\}$. The observed sample mean in the $g$th group is $\overline{\mathcal{C}}_{g} = \frac{1}{n_{g}} \sum_{i=1}^{n_{g}} \mathcal{C}_{gi}$ and the total collection of sample means $\overline{\mathcal{C}}(N) = \{\overline{\mathcal{C}}_{1}, \ldots, \overline{\mathcal{C}}_{g_{0}}\}$. 

We thus consider the adjustment of $\langle \mathcal{M}(\mathcal{C}) \rangle$ given $\mathcal{C}(N)$. By exploiting the concept of Bayes linear sufficiency (Goldstein and O'Hagan, 1996)\nocite{Goldstein/O'Hagan:1996} the following lemma shows that we can restrict attention to the observation of the sample means $\overline{\mathcal{C}}(N)$ rather than $\mathcal{C}(N)$; the proof is in the appendix.
\begin{Lem} \label{lem:mean} The collection $\overline{\mathcal{C}}(N)$ is Bayes linear sufficient for $\mathcal{C}(N)$ for adjusting $\mathcal{M}(\mathcal{C})$. Thus, as $\langle \overline{\mathcal{C}}(N) \rangle \subseteq \langle \mathcal{C}(N) \rangle$, we have $E_{\mathcal{C}(N)}(\mathcal{M}(\mathcal{C})) = E_{\overline{\mathcal{C}}(N)}(\mathcal{M}(\mathcal{C}))$ and $Var_{\mathcal{C}(N)}(\mathcal{M}(\mathcal{C})) = Var_{\overline{\mathcal{C}}(N)}(\mathcal{M}(\mathcal{C}))$.
\end{Lem}
Hence, the adjustment of $\langle \mathcal{M}(\mathcal{C}) \rangle$ given $\mathcal{C}(N)$ is identical to the adjustment of $\langle \mathcal{M}(\mathcal{C}) \rangle$ given $\overline{\mathcal{C}}(N)$. As in Shaw and Goldstein (1999)\nocite{Shaw/Goldstein:1999} we will show that this adjustment can be performed by considering separately the analysis of variables and of groups where each analysis has an interpretable form. In the latter analysis, for both insight and to enable us to draw a close parallel with the finite work we develop in Section \ref{sec:finite}, we will use Bayes linear sufficiency to exploit the generalised conditional independence structure (Smith, 1990)\nocite{Smith:1990} of the model.
\subsection{Variable analysis: adjustment of  $\langle \mathcal{M}(\mathcal{C}_{g}) \rangle$ given $\mathcal{C}_{g}(n_{g})$ \label{sec:variables}}
We consider, for each $g = 1, \ldots, g_{0}$, the adjustment of quantities contained in $\langle \mathcal{M}(\mathcal{C}_{g}) \rangle$ given $\mathcal{C}_{g}(n_{g})$ so that we sample $n_{g}$ individuals from an infinite second-order exchangeable population and adjust the underlying population mean vector. We will show that this variable problem requires only the solution of a single $v_{0} \times v_{0}$ generalised eigenvalue problem involving the matrices $D$ and $C$ defined in Assumption \ref{assum2}.
\begin{Def} \label{def:var}
The underlying canonical variable directions are defined as the columns of the matrix $U = [U_{1} \ldots U_{v_{0}}]$ solving the generalised eigenvalue problem $CU = DU\Phi$, where $\Phi = diag(\phi_{1}, \ldots, \phi_{v_{0}})$ is the matrix of eigenvalues. $U$ is chosen so that $U^{T}CU = I$, $U^{T}DU\Phi = I$. The ordered eigenvalues $1 > \phi_{1} \geq \cdots \geq \phi_{v_{0}} > 0$ are termed the underlying canonical variable resolutions.
\end{Def}
Let $U_{vt}$ denote the $v$th component of the $t$th canonical variable direction and, for each $t=1, \ldots, v_{0}$, define $W_{t} \in \langle \mathcal{C} \rangle$ to be $W_{t} = \sum_{v=1}^{v_{0}} U_{vt}X_{v}$. For each $g = 1, \ldots, g_{0}$ create $\mathcal{M}(W_{gt}) = \sum_{v=1}^{v_{0}} U_{vt}\mathcal{M}(X_{gv})$, and for each $i$, $W_{gti} = \sum_{v=1}^{v_{0}} U_{vt}X_{gvi}$ with $\overline{W}_{gt} = \frac{1}{n_{g}} \sum_{i=1}^{n_{g}} W_{gti}$. Hence, $W_{t}$, $\mathcal{M}(W_{gt})$, $W_{gti}$ and $\overline{W}_{gt}$ share the same coordinate representation. From Theorem 1 of Shaw and Goldstein (2012)\nocite{Shaw/Goldstein:2012}, a reformulated version of Theorem 3 of Goldstein and Wooff (1998)\nocite{Goldstein/Wooff:1998}, we have the following corollary.
\begin{Cor} \label{cor:var}
For a sample of size $n_{g}$, $\mathcal{C}_{g}(n_{g})$, drawn from an infinite population, the collection $\mathcal{M}(\mathcal{W}^{*}_{g}) = \{\mathcal{M}(W_{g1}), \ldots, \mathcal{M}(W_{gv_{0}})\}$ forms a basis for $\lincom{\mathcal{M}(\mathcal{C}_{g})}$. The $\mathcal{M}(W_{gt})$ are a priori uncorrelated, and, for all samples of any size, a posteriori uncorrelated and are the canonical directions for the adjustment. The posterior adjusted expectation, $\mu_{[g: n_{g}]t} = E_{\mathcal{C}_{g}(n_{g})}(\mathcal{M}(W_{gt}))$, and posterior adjusted precision, $r_{[g: n_{g}]t} = Var^{-1}_{\mathcal{C}_{g}(n_{g})}(\mathcal{M}(W_{gt}))$, are given by
\begin{eqnarray}
\mu_{[g: n_{g}]t} \ = \ \frac{r_{[g:0]t}\mu_{[g:0]t}+n_{g}r_{[g]t}\overline{W}_{gt}}{r_{[g:0]t} + n_{g}r_{[g]t}}; \ r_{[g: n_{g}]t} \ = \  r_{[g:0]t} + n_{g}r_{[g]t}\label{eq:var}
\end{eqnarray}
where $\mu_{[g:0]t} = E(\mathcal{M}(W_{gt}))$, $r_{[g:0]t} = Var^{-1}(\mathcal{M}(W_{gt}))$ and $r_{[g]t} = Var^{-1}_{\mathcal{M}(\mathcal{C}_{g})}(W_{gti})$.
\end{Cor}
If appropriately scaled to have a prior variance of one, the $\mathcal{M}(\mathcal{W}^{*}_{g})$ form an orthonormal grid over $\lincom{\mathcal{M}(\mathcal{C}_{g})}$ which summarises the effect of the sample. The coordinate representation of the $\mathcal{M}(\mathcal{W}^{*}_{g})$ is the same for each group $g$ and for each sample size $n_{g}$. Therefore, the qualitative features of the update are also the same for each group and for each sample size which makes it straightforward to assess the affect of differing sample sizes both within and across groups. As Shaw and Goldstein (2012)\nocite{Shaw/Goldstein:2012} note, the corresponding quantitative features generalise the familiar univariate normal model results to individual quantities in $\langle \mathcal{M}(\mathcal{C}_{g}) \rangle$. The posterior adjusted precision is the sum of the prior precision and the data precision, the latter being the product of the sample size and the data precision for a single observation. The posterior adjusted expectation is the weighted average of the prior expectation and the sample mean, weighted according to the corresponding precisions. 

If we let $\lambda_{[g: n_{g}]t}$ denote the resolution of $\mathcal{M}(W_{gs})$ given $\mathcal{C}_{g}(n_{g})$ we have
\begin{eqnarray}
\lambda_{[g: n_{g}]t} & = & \frac{n_{g}\alpha_{gg} \phi_{t}}{(n_{g}-1) \alpha_{gg} \phi_{t} + \gamma_{g}}. \label{eq:res}
\end{eqnarray}
We thus observe the role of the $\phi_{t}$s, the underlying canonical variable resolutions, and also how to amend them to account for the group factors and sample size when obtaining $\lambda_{[g: n_{g}]t}$. Notice that, for all $g$ and $t \neq s$, $\lambda_{[g: n_{g}]t} > \lambda_{[g: n_{g}]s}$ if and only if $\phi_{t} > \phi_{s}$. As $\lambda_{[g: n_{g}]t}$ is an increasing function of $n_{g}$ then the numerical order of the canonical resolutions is the same for each group and sample size and is that given by the ordering of the $\phi_{t}$. Goldstein and Wooff (1998, Corollary 1)\nocite{Goldstein/Wooff:1998} show how the simplicity of the dependence of the sample size on the resolution when performing second-order exchangeable sampling can be exploited to simplify any design problem which requires a sample size to be chosen to obtain a given variance reduction over quantities contained in $\langle \mathcal{M}(\mathcal{C}_{g}) \rangle$. Equally, we can use either of equations (\ref{eq:var}) and (\ref{eq:res}) to compare variance reductions in differing groups. For example, for any $g$, $h$ we have
\begin{eqnarray}
\lambda_{[g: n_{g}]t} >  \lambda_{[h: n_{h}]t} \ \Leftrightarrow \ \frac{n_{g}r_{[g]t}}{r_{[g:0]t}} > \frac{n_{h}r_{[h]t}}{r_{[h:0]t}} \ \Leftrightarrow \ \left(\frac{\gamma_{h}}{n_{h}\alpha_{hh}} - \frac{\gamma_{hg}}{n_{g}\alpha_{gg}}\right)\phi_{t}^{-1} - \left(\frac{1}{n_{h}} - \frac{1}{n_{g}}\right) > 0. \label{eq:varres}
\end{eqnarray}
Thus, for each $\mathcal{M}(W_{gt})$, we learn most in the groups with the highest ratio of prior variance (of $\mathcal{M}(W_{gt})$) to likelihood variance (of $\overline{W}_{gt}$). If we observe the same sample size in each group then, for every $t$, we learn most in the groups with high values of $\alpha_{gg}/\gamma_{g}$. Observe the computational advantage that, irrespective of the number of groups or the number of observations, we can obtain the results of Corollary \ref{cor:var} for each group through the solution of a single $v_{0} \times v_{0}$ generalised eigenvalue problem, that given in Definition \ref{def:var}. 

From the specifications given by Assumptions \ref{assum1} and \ref{assum2}, $Var(\mathcal{M}(W_{gt}))$ is separable between $g$ and $t$ whereas, in general, $Var_{\mathcal{M}(\mathcal{C}_{g})}(W_{gti})$ is not. If it is then each precision term in Corollary \ref{cor:var} becomes separable between $g$ and $t$ as Corollary \ref{cor:var1} demonstrates. These results are exactly those that would have been obtained had we directly utilised the model of Shaw and Goldstein (1999)\nocite{Shaw/Goldstein:1999}, see Section \ref{sec:3.1.2}.
\begin{Cor} \label{cor:var1}
If, for all $g = 1, \ldots, g_{0}$, $\alpha_{gg}/\gamma_{g} = a$ for some constant $\phi_{1}^{-1} > a > 0$ then, for the precisions defined in Corollary \ref{cor:var}, $r_{[g: 0]t} = \alpha_{gg}^{-1}r_{0t}$ and $r_{[g]t} = \gamma^{-1}_{g}r_{t}$ where both $r_{0t} = Var^{-1}(\alpha_{gg}^{-\frac{1}{2}}\mathcal{M}(W_{gt}))$ and, for any $i$, $r_{t} = Var^{-1}_{\mathcal{M}(\mathcal{C}_{g})}(\gamma_{g}^{-\frac{1}{2}}W_{gti})$ do not depend upon $g$.
\end{Cor}
\subsection{Variable analysis for the examination data example \label{sec:exvar}}
The examiner uses his specifications given in Section \ref{sec:ex1} to find the corresponding underlying canonical variable directions and resolutions. The latter are 
\begin{eqnarray}
\phi_{1} = \cdots = \phi_{4} = 0.8, \ \phi_{5} = \phi_{6} = 0.4778, \ \phi_{7} = 0.1666, \ \phi_{8} = 0.1133. \label{eq:phis}
\end{eqnarray}
 The four directions paired with $\phi_{1}, \ldots, \phi_{4}$ correspond to any four mutually uncorrelated linear combinations $\mathcal{M}(W_{gt}) = \sum_{v=1}^{5} U_{vt}\mathcal{M}(X_{gv})$ with $\sum_{v=1}^{5} U_{vt} = 0$ and so represent differences between the difficulties of Section A questions under marker $g$. Choosing the cumulative residuals, that is the contrasts derived from the Helmert matrix of order five, we have
\begin{eqnarray}
\mathcal{M}(W_{gt}) & \propto & \mathcal{M}(X_{gt+1}) - \frac{1}{t}\sum_{v=1}^{t} \mathcal{M}(X_{gv}), \ t = 1, \ldots, 4. \label{eq:secAdiff}
\end{eqnarray}
Similarly, the two directions paired with $\phi_{5}$ and $\phi_{6}$ correspond to any two mutually uncorrelated linear combinations $\mathcal{M}(W_{gv}) = \sum_{v=6}^{8} U_{vt}\mathcal{M}(X_{gv})$ with $\sum_{v=6}^{8} U_{vt} = 0$ and represent differences between the difficulties of Section B questions under marker $g$. Once again, choosing the cumulative residuals, we have $\mathcal{M}(W_{g5}) \propto \mathcal{M}(X_{g7}) - \mathcal{M}(X_{g6})$ and $\mathcal{M}(W_{g6}) \propto \mathcal{M}(X_{g8}) - 0.5\mathcal{M}(X_{g7}) - 0.5\mathcal{M}(X_{g6})$. The remaining two directions, paired with $\phi_{7}$ and $\phi_{8}$ respectively, enable us to compare the differences and similarities between the two sections of the examination. We have $\mathcal{M}(W_{g7}) = 0.2683TotA_{g}+0.0702TotB_{g}$ and $\mathcal{M}(W_{g8}) = 0.3572TotA_{g}-0.3499TotB_{g}$ where, as in Section \ref{sec:ex1}, $TotA_{g}$, $TotB_{g}$ are the underlying script totals on Sections A and B for the $g$th marker. Thus, we can use linear combinations of $\mathcal{M}(W_{g7})$ and $\mathcal{M}(W_{g8})$ to look at the relationships between the two sections of the examination. For example, the underlying total score $Tot_{g} = TotA_{g} + TotB_{g} = 5.9437\mathcal{M}(W_{g7})-1.6649\mathcal{M}(W_{g8})$ and the difference between the section totals, $TotA_{g} - TotB_{g} = -0.0610\mathcal{M}(W_{g7})+2.8454\mathcal{M}(W_{g8})$. We can thus see the value of the resolution partition, see equation (\ref{eq:respart}), in this case. For a sample of $n_{g}$ individuals in the $g$th group, $Res_{[g:n_{g}]}(Tot_{g}) = 0.9272\lambda_{[g: n_{g}]7}+0.0728\lambda_{[g: n_{g}]8}$ and $Res_{[g:n_{g}]}(TotA_{g} - TotB_{g}) = 0.0005\lambda_{[g: n_{g}]7}+0.9995\lambda_{[g: n_{g}]8}$ where $\lambda_{[g: n_{g}]t}$ is the corresponding resolution of $\mathcal{M}(W_{gt})$. Thus, most of the information about the total on the script is carried by $\mathcal{M}(W_{g7})$ whilst $\mathcal{M}(W_{g8})$ is highly informative about the difference in total performance on the two sections.

Notice that, as the examiner specified $\gamma_{g} = \alpha_{gg} = 1$ for all $g$ then, from (\ref{eq:res}), $\lambda_{[g: n_{g}]t} = n_{g}\phi_{t}/\{(n_{g}-1)\phi_{t}+1\}$ which reflects the additional symmetry over groups imposed by the examiner's beliefs: if we observe a sample of size $n$ in the $g$th group then not only do the canonical directions share the same coordinate representation for each $g$ but the corresponding canonical resolutions are identical. 

\subsection{Group analysis: adjustment of  $\langle \mathcal{M}(\mathcal{W}_{t}) \rangle$ given $\overline{\mathcal{W}}_{t}$ \label{sec:groups}}
We now consider adjustment across different groups. For each $t = 1, \ldots, v_{0}$ form the collections $\mathcal{M}(\mathcal{W}_{t}) = \{\mathcal{M}(W_{1t}), \ldots, \mathcal{M}(W_{g_{0}t})\}$ and $\overline{\mathcal{W}}_{t} = \{\overline{W}_{1t}, \ldots, \overline{W}_{g_{0}t}\}$. Thus, we form $\mathcal{M}(\mathcal{W}_{t})$ by collecting together the $t$th canonical variable direction from each of the $g_{0}$ groups and $\overline{\mathcal{W}}_{t}$ is the collection of sample means from each group which share the same coordinate representation as each $\mathcal{M}(W_{gt})$. We also form $\mathcal{M}(\mathcal{W}) = \{\mathcal{M}(\mathcal{W}_{1}), \ldots, \mathcal{M}(\mathcal{W}_{v_{0}})\}$, the total collection of canonical variable directions. In Section \ref{sec:combined} we will show that by considering, for each $t$, the adjustment of $\langle \mathcal{M}(\mathcal{W}_{t}) \rangle$ given $\overline{\mathcal{W}}_{t}$ we can completely determine the adjustment of $\langle \mathcal{M}(\mathcal{C}) \rangle$ given $\mathcal{C}(N)$. Having established, in Lemma \ref{lem:mean}, that $\overline{\mathcal{C}}(N)$ is Bayes linear sufficient for $\mathcal{C}(N)$ for adjusting $\mathcal{M}(\mathcal{C})$ in the following lemma we show that for the adjustment of $\langle \mathcal{M}(\mathcal{W}_{t}) \rangle$ we may restrict attention to $\overline{\mathcal{W}}_{t}$ rather than $\overline{\mathcal{C}}(N)$; the proof is in the appendix.
\begin{Lem} \label{lem:meansuff}
For each $t = 1, \ldots, v_{0}$, the collection $\overline{\mathcal{W}}_{t}$ is Bayes linear sufficient for $\overline{\mathcal{C}}(N)$ for adjusting $\mathcal{M}(\mathcal{W}_{t})$ so that, as $\langle \overline{\mathcal{W}}_{t} \rangle \subseteq \langle \overline{\mathcal{C}}(N) \rangle$, $E_{\overline{\mathcal{C}}(N)}(\mathcal{M}(\mathcal{W}_{t})) = E_{\overline{\mathcal{W}}_{t}}(\mathcal{M}(\mathcal{W}_{t}))$ and $Var_{\overline{\mathcal{C}}(N)}(\mathcal{M}(\mathcal{W}_{t})) = Var_{\overline{\mathcal{W}}_{t}}(\mathcal{M}(\mathcal{W}_{t}))$.
\end{Lem}
We now consider, for each $t$, the adjustment of $\langle \mathcal{M}(\mathcal{W}_{t}) \rangle$ given $\overline{\mathcal{W}}_{t}$. The solution can be obtained through the solution of a $g_{0} \times g_{0}$ generalised eigenvalue problem involving the matrices $A$, $\hat{A}$ and $B$ defined in Assumption \ref{assum2} and $N$, the diagonal matrix of sample sizes.
\begin{Def} \label{def:group}
For each $t = 1, \ldots, v_{0}$, the $(s, t)$th canonical group direction is defined to be the $s$th column of the matrix $V_{t} = [V_{1t} \ldots V_{g_{0}t}]$ solving the generalised eigenvalue problem $AV_{t} = (A + \phi_{t}^{-1}N^{-1}B - N^{-1}\hat{A})V_{t}\Lambda_{t}$, where $\phi_{t}$ is the $t$th underlying canonical variable resolution and $\Lambda_{t} = diag(\lambda_{1t}, \ldots, \lambda_{g_{0}t})$ is the matrix of eigenvalues ordered so that $1 > \lambda_{1t} \geq \cdots \geq \lambda_{g_{0}t} > 0$. $\lambda_{st}$ is termed the $(s, t)$th canonical group resolution. $V_{t}$ is chosen so that $V_{t}^{T}AV_{t} = I$ and $V_{t}^{T}(A + \phi_{t}^{-1}N^{-1}B-N^{-1}\hat{A})V_{t}\Lambda_{t} = I$.
\end{Def}
Let $V_{gst}$ denote the $g$th component of the $(s, t)$th canonical group direction and, for each $s=1, \ldots, g_{0}$, create $\mathcal{M}(Y_{st}) = \sum_{g=1}^{g_{0}} V_{gst}\mathcal{M}(W_{gt})$ and $\overline{Y}_{st}  = \sum_{g=1}^{g_{0}} V_{gst}\overline{W}_{gt}$. We have the following theorem; the proof is in the appendix.
\begin{Theo} \label{theo:group} 
For each $t =1, \ldots, v_{0}$, the collection $\mathcal{M}(\mathcal{Y}_{t}) = \{\mathcal{M}(Y_{1t}), \ldots, \mathcal{M}(Y_{g_{0}t})\}$ forms a basis for $\langle \mathcal{M}(\mathcal{W}_{t}) \rangle$. The $\mathcal{M}(Y_{st})$ are a priori uncorrelated, and, given $\overline{\mathcal{W}}_{t}$, a posteriori uncorrelated and are the canonical directions for the adjustment of $\langle \mathcal{M}(\mathcal{W}_{t}) \rangle$. The posterior adjusted expectation, $\mu_{Nst} = E_{\overline{\mathcal{W}}_{t}}(\mathcal{M}(Y_{st}))$, and posterior adjusted precision, $r_{Nst} = Var^{-1}_{\overline{\mathcal{W}}_{t}}(\mathcal{M}(Y_{st}))$, are given by
\begin{eqnarray}
\mu_{Nst} \ = \ \frac{r_{0st}\mu_{0st} + r_{st}\overline{Y}_{st}}{r_{0st}+r_{st}}; \ r_{Nst} \ = \ r_{0st} + r_{st}, \label{eq:group}
\end{eqnarray}
where $\mu_{0st} = E(\mathcal{M}(Y_{st}))$, $r_{0st} = Var^{-1}(\mathcal{M}(Y_{st}))$ and $r_{st} = Var^{-1}_{\mathcal{M}(\mathcal{C})}(\overline{Y}_{st})$. The resolution of $\mathcal{M}(Y_{st})$ given $\overline{\mathcal{W}}_{t}$ is $\lambda_{st}$.
\end{Theo}
Theorem \ref{theo:group} provides us with the wherewithal to perform a canonical analysis for the adjustment of $\langle \mathcal{M}(\mathcal{W}_{t}) \rangle$ given $\overline{\mathcal{W}}_{t}$. For example, see equation (\ref{eq:respart}), we have the resolution partition that for any $\mathcal{M}(Z_{t}) = \sum_{g=1}^{g_{0}} \eta_{g} \mathcal{M}(W_{gt}) \in \langle \mathcal{M}(\mathcal{W}_{t}) \rangle$, $Res_{\overline{\mathcal{W}}_{t}}(\mathcal{M}(Z_{t})) = \sum_{s=1}^{g_{0}} Corr^{2}(\mathcal{M}(Z_{t}),$ $\mathcal{M}(Y_{st}))\lambda_{st}$ where $\sum_{s=1}^{g_{0}} Corr^{2}(\mathcal{M}(Z_{t}), \mathcal{M}(Y_{st})) = 1$. Note, see the proof to Theorem \ref{theo:group}, that whilst $Var(\mathcal{M}(Y_{st}))$ is separable between $s$ and $t$, $Var_{\mathcal{M}(\mathcal{C})}(\overline{Y}_{st})$ is not. The coordinate representation of each $\mathcal{M}(Y_{st})$ will typically, via $\phi_{t}$, depend upon each $t$ and the corresponding $\lambda_{st}$ will not be a straightforward function of $\phi_{t}$. In general, we will be required to solve separably each of the $v_{0}$ $g_{0} \times g_{0}$ generalised eigenvalue problems in Definition \ref{def:group}. However, in certain special cases we can extract for each $t$, $\mathcal{M}(\mathcal{Y}_{t})$ and $\Lambda_{t}$ from the solution of a single $g_{0} \times g_{0}$ generalised eigenvalue problem as the following corollary shows.
\begin{Cor} \label{cor:prop}
If for all $g = 1, \ldots, g_{0}$, $\alpha_{gg}/\gamma_{g} = a$ for some constant $\phi_{1}^{-1} > a > 0$ then, for the precisions defined in Theorem \ref{theo:group}, $r_{0st} = q_{0s}r_{0t}$ and $r_{st} = q_{s}r_{t}$ where both $q_{0s} = Var^{-1}(r_{0t}^{\frac{1}{2}}\mathcal{M}(Y_{st}))$ and $q_{s} = Var^{-1}_{\mathcal{M}(\mathcal{C})}(r_{t}^{\frac{1}{2}}\overline{Y}_{st})$ do not depend upon $t$ and $r_{0t}$ and $r_{t}$ are as defined in Corollary \ref{cor:var1}. The $(s, t)$th canonical group direction is the same for each $t$ and is the $s$th column of the matrix $V = [V_{1} \, \ldots \, V_{g_{0}}]$ solving the generalised eigenvalue problem $AV = (A + N^{-1}B)V\Psi$ where $V$ is chosen so that $V^{T}AV = I$ and $V^{T}(A + N^{-1}B)V\Psi = I$ and $\Psi = diag(\psi_{1}, \ldots, \psi_{g_{0}})$ is the matrix of eigenvalues ordered so that $1 > \psi_{1} \geq \cdots \geq \psi_{g_{0}} > 0$. The $(s, t)$th canonical group resolution is given by
\begin{eqnarray}
\lambda_{st} & = & \frac{\psi_{s}\phi_{t}}{\psi_{s}\phi_{t} + (1-\psi_{s})(1-a\phi_{t})}. \label{eq:separable}
\end{eqnarray}
\end{Cor}
 The results of Corollary \ref{cor:prop} coincide with those obtained using the model of Shaw and Goldstein (1999)\nocite{Shaw/Goldstein:1999} though an alternate motivation for the group analysis is given in that paper.

Notice that, in contrast to the adjustment of $\langle \mathcal{M}(\mathcal{C}_{g}) \rangle$ given $\mathcal{C}_{g}(n_{g})$, comparing possible choices of sample size for the adjustment of $\langle \mathcal{M}(\mathcal{W}_{t}) \rangle$ given $\overline{\mathcal{W}}_{t}$ will not be straightforward. $\mathcal{M}(\mathcal{Y}_{t})$ and $\Lambda_{t}$ will, see Definition \ref{def:group}, typically depend upon the individual group sample sizes $n_{1}$, \ldots, $n_{g_{0}}$ in a less than tractable way. However, in the balanced design, with $n_{g} = n$ for all $g$, the adjustment of $\langle \mathcal{M}(\mathcal{W}_{t}) \rangle$ given $\overline{\mathcal{W}}_{t}$ can be viewed as sampling from a infinite second-order exchangeable population. Thus, $\mathcal{M}(\mathcal{Y}_{t})$ will not depend upon $n$ and $\Lambda_{t}$ will be a simple function of $n$. To see this, for each $t=1, \ldots, v_{0}$ and each $i$ form the collection $\mathcal{W}_{ti} = \{W_{1ti}, \ldots, W_{g_{0}ti}\}$ so that $\mathcal{W}_{ti}$ contains a single individual from each group and we use the labelling convention that this is the $i$th individual. From Assumptions \ref{assum1} and \ref{assum2} and Definition \ref{def:var} it can easily be verified that the $\mathcal{W}_{ti}$ are infinitely second-order exchangeable over $i$ and so, using the second-order representation theorem of Goldstein (1986)\nocite{Goldstein:1986}, we can write $\mathcal{W}_{ti} = \mathcal{M}(\mathcal{W}_{t}) + \mathcal{R}_{i}(\mathcal{W}_{t})$. If we observe a sample of size $n$, $\mathcal{W}_{t1}, \ldots, \mathcal{W}_{tn}$, then the sample mean $\overline{\mathcal{W}}_{t}$ is Bayes linear sufficient for the adjustment of $\langle \mathcal{M}(\mathcal{W}_{t}) \rangle$. Hence, in the balanced design, the adjustment of $\langle \mathcal{M}(\mathcal{W}_{t}) \rangle$ given $\overline{\mathcal{W}}_{t}$, as described in Theorem \ref{theo:group}, is precisely one of sampling $n$ individuals from an infinitely exchangeable second-order population and so using Theorem 1 of Shaw and Goldstein (2012)\nocite{Shaw/Goldstein:2012}, a reformulated version of Theorem 3 of Goldstein and Wooff (1998), we have the following corollary.
\begin{Cor} \label{cor:balanced} If $N= nI$ then the collection $\mathcal{M}(\mathcal{Y}_{t})$, as defined in Theorem \ref{theo:group}, does not depend upon $n$. The posterior adjusted expectation $\mu_{Nst}$, posterior adjusted precision $r_{Nst}$ and resolution $\lambda_{st}$ of each $\mathcal{M}(Y_{st})$ given $\overline{\mathcal{W}}_{t}$ are given by
\begin{eqnarray*} 
\mu_{Nst} \ = \ \frac{r_{0st}\mu_{0st} + nq_{st}\overline{Y}_{st}}{r_{0st}+nq_{st}}; \ r_{Nst} \ = \ r_{0st} + nq_{st}; \ \lambda_{st} \ = \ \frac{n \psi_{st}}{(n-1)\psi_{st} + 1}
\end{eqnarray*}
where $q_{st} = Var^{-1}_{\mathcal{M}(\mathcal{W}_{t})}(Y_{sti})$, $Y_{sti} = \sum_{g=1}^{g_{0}} V_{gst}W_{gti}$ shares the same coordinate representation as $\mathcal{M}(Y_{st})$ and $\psi_{st}$ is the $s$th eigenvalue solving $AV_{t} = (A + \phi_{t}^{-1}B - \hat{A})V_{t}\Psi_{t}$.
\end{Cor}
Note that we can combine the results of Corollaries \ref{cor:prop} and \ref{cor:balanced} to observe that, in the balanced design with $\alpha_{gg}/\gamma_{g} = a$, for all choices of $n$ and $t$ we can extract each $\mathcal{M}(Y_{st})$ and $\lambda_{st}$ from a single $g_{0} \times g_{0}$ eigenvalue problem: $AV = (A + B)V\Psi_{(1)}$ where $\Psi_{(1)} = diag(\psi_{(1)1}, \ldots, \psi_{(1)g_{0}})$. Thus, in Corollary \ref{cor:prop}, $\psi_{s}$ and $\lambda_{st}$ are respectively given by
\begin{eqnarray}
\psi_{s} \ = \ \frac{n\psi_{(1)s}}{(n-1)\psi_{(1)s} + 1}; \ \lambda_{st} \ = \ \frac{n\psi_{(1)s}\phi_{t}}{n\psi_{(1)s}\phi_{t} + (1-\psi_{(1)s})(1-a\phi_{t})}. \label{eq:sepbalanced}
\end{eqnarray}
This result coincides with Corollary 1 of Shaw and Goldstein (1999)\nocite{Shaw/Goldstein:1999}.
\subsection{Variable and group analyses combined: adjustment of  $\langle \mathcal{M}(\mathcal{C}) \rangle$ given $\mathcal{C}(N)$ \label{sec:combined}}
We now show that the canonical variable analysis of Section \ref{sec:variables} and canonical group analysis of Section \ref{sec:groups} completely determines the posterior analysis of the collection $\mathcal{M}(\mathcal{C})$ given $\mathcal{C}(N)$. To do this we exploit Bayes linear sufficiency as described in the following lemma; the proof is in the appendix.
\begin{Lem} \label{lem:sep}
1. The collection $\mathcal{M}(\mathcal{W})$ is Bayes linear sufficient for $\overline{\mathcal{C}}(N)$ for adjusting $\mathcal{M}(\mathcal{C})$ so that
\begin{eqnarray*}
E_{\overline{\mathcal{C}}(N)}(\mathcal{M}(\mathcal{C})) & = & E_{\overline{\mathcal{C}}(N)}\{E_{\mathcal{M}(\mathcal{W})}(\mathcal{M}(\mathcal{C}))\}; \\
Var_{\overline{\mathcal{C}}(N)}(\mathcal{M}(\mathcal{C})) & = & Var_{\mathcal{M}(\mathcal{W})}(\mathcal{M}(\mathcal{C})) +  Var_{\overline{\mathcal{C}}(N)}\{E_{\mathcal{M}(\mathcal{W})}(\mathcal{M}(\mathcal{C}))\}.
\end{eqnarray*}

2. For all $t \neq t'$, the collection $\overline{\mathcal{C}}(N)$ is Bayes linear sufficient for $\mathcal{M}(\mathcal{W}_{t'})$ for adjusting $\mathcal{M}(\mathcal{W}_{t})$ so that $Cov_{\overline{\mathcal{C}}(N)}(\mathcal{M}(\mathcal{W}_{t}), \mathcal{M}(\mathcal{W}_{t'})) = 0$.
\end{Lem}
Note that for any $\mathcal{M}(Z) \in \langle \mathcal{M}(\mathcal{C}) \rangle$ we have $\mathcal{M}(Z) = \sum_{s=1}^{g_{0}} \sum_{t=1}^{v_{0}} a_{st} \mathcal{M}(Y_{st}) \in \langle \mathcal{M}(\mathcal{W}) \rangle$ so that $E_{\mathcal{M}(\mathcal{W})}(\mathcal{M}(Z)) = \mathcal{M}(Z)$ and $Var_{\mathcal{M}(\mathcal{W})}(\mathcal{M}(Z)) = 0$. Consequently, using the first part of Lemma \ref{lem:sep}, for the adjustment of $\langle \mathcal{M}(\mathcal{C}) \rangle$ given $\overline{\mathcal{C}}(N)$ we only need consider the adjustment of $\langle \mathcal{M}(\mathcal{W}) \rangle$ given $\overline{\mathcal{C}}(N)$. From the second part of Lemma \ref{lem:sep}, we may deduce that $Var_{\overline{\mathcal{C}}(N)}(\mathcal{M}(\mathcal{W})) = \oplus_{t=1}^{v_{0}} Var_{\overline{\mathcal{C}}(N)}(\mathcal{M}(\mathcal{W}_{t}))$ where $\oplus$ denotes the direct sum. Further, as $Cov(\mathcal{M}(\mathcal{W}_{t}), \mathcal{M}(\mathcal{W}_{t'})) = 0$ for $t \neq t'$ then $T_{\mathcal{M}(\mathcal{W}) \, : \, \overline{\mathcal{C}}(N)} = \oplus_{t=1}^{v_{0}} T_{\mathcal{M}(\mathcal{W}_{t}) \, : \, \overline{\mathcal{C}}(N)}$. Hence, additionally using Lemmas \ref{lem:mean} and \ref{lem:meansuff}, we have the following corollary to Theorem \ref{theo:group}.
\begin{Cor} \label{cor:full}
The collection $\mathcal{M}(\mathcal{Y}) = \{\mathcal{M}(Y_{st}), s=1, \ldots, g_{0}, t = 1, \ldots, v_{0}\}$ forms a basis for $\langle \mathcal{M}(\mathcal{C}) \rangle$. The $\mathcal{M}(Y_{st})$ are a priori uncorrelated, and, given $\mathcal{C}(N)$, a posteriori uncorrelated and are the canonical directions for the adjustment of $\langle \mathcal{M}(\mathcal{C}) \rangle$ given $\mathcal{C}(N)$ with corresponding canonical resolutions given by the $\lambda_{st}$.
\end{Cor}
Thus, we have the computational simplicity that the canonical analysis of the $g_{0}v_{0} \times g_{0}v_{0}$ problem may be obtained from the canonical analysis of a single $v_{0} \times v_{0}$ problem, as given by Definition \ref{def:var}, which corresponds to sampling from an infinite second-order exchangeable population, and the canonical analysis of $v_{0}$ related $g_{0} \times g_{0}$ problems, as given by Definition \ref{def:group}. Only the $g_{0} \times g_{0}$ problems depend upon the sample sizes and, in the case of the balanced design, correspond to sampling from an infinite second-order exchangeable population.

\subsection{Group and combined analysis for the examination data example\label{sec:exinf}}
The examiner elects to take a sample of $n$ scripts from each of the three markers and considers the adjustment of $\langle \mathcal{M}(\mathcal{C}) \rangle$ given $\mathcal{C}(N)$ where $N = diag(n, n, n)$ in order to, see Section \ref{sec:ex1}, answer his questions of interest.  From Corollary \ref{cor:full}, he can obtain the desired canonical structure of this adjustment by, see Theorem \ref{theo:group}, considering a group analysis over the underlying canonical variable directions he found in Section \ref{sec:exvar}. He thus forms, for each $t= 1, \ldots, 8$, the collections $\mathcal{M}(\mathcal{W}_{t}) = \{\mathcal{M}(W_{1t}), \ldots, \mathcal{M}(W_{3t})\}$ and $\overline{\mathcal{W}}_{t} = \{\overline{W}_{1t}, \ldots, \overline{W}_{3t}\}$. Hence, for example, as $\mathcal{M}(W_{g5}) \propto \mathcal{M}(X_{g7}) - \mathcal{M}(X_{g6})$ then $\mathcal{M}(\mathcal{W}_{5})$ represents the collection, across the three markers, of the underlying mean difference between the score on question seven and that on question six whilst $\overline{\mathcal{W}}_{5}$ is the collection of observed sample average difference of the scores on these questions.

In general, for each set of sample sizes $N$, the examiner has to solve the eight $3 \times 3$ generalised eigenvalue problems given in Definition \ref{def:group}. However, in this case, the collection $\mathcal{M}(\mathcal{Y}_{t})$ does not depend upon either $t$ (via Corollary \ref{cor:prop} as $\gamma_{g} = \alpha_{gg} = 1$ for all $g$) or sample size (via Corollary \ref{cor:balanced}  as the design is balanced). Thus, for any $n$, only the solution of the $3 \times 3$ problem $AV = (A+B)V\Psi_{(1)}$ is needed.  The examiner finds that $\psi_{(1)1} = 27/37$ and $\psi_{(1)2} = \psi_{(1)3} = 3/23$ and thus, from (\ref{eq:sepbalanced}), for each $t$ $\lambda_{1t} = 27\phi_{t}n/\{27\phi_{t}n + 10(1-\phi_{t})\}$ and $\lambda_{2t} = \lambda_{3t} = 3\phi_{t}n/\{3\phi_{t}n + 20(1-\phi_{t})\}$ so that there are eight distinct $\lambda_{st}$. The matrix $V$ can be obtained from the appropriately scaled Helmert matrix of order three so that $\mathcal{M}(Y_{1t}) \propto \mathcal{M}(W_{1t}) + \mathcal{M}(W_{2t}) + \mathcal{M}(W_{3t})$ whilst $\mathcal{M}(Y_{2t}) \propto  \mathcal{M}(W_{1t}) - \mathcal{M}(W_{2t})$ and $\mathcal{M}(Y_{3t}) \propto \mathcal{M}(W_{1t}) - 0.5 \mathcal{M}(W_{2t}) -0.5 \mathcal{M}(W_{3t})$. Thus, $\mathcal{M}(Y_{1t})$ is the total (or average) across the groups of the $t$th canonical variable directions whilst $\mathcal{M}(Y_{2t})$ and $\mathcal{M}(Y_{3t})$ are two linear contrasts. 

The symmetry of the beliefs made by the examiner result in many of his questions of interest relating either directly to the $\mathcal{M}(Y_{st})$ or being closely correlated to them. For example for $t = 1, \ldots, 4$, see (\ref{eq:secAdiff}), $\mathcal{M}(Y_{1t})$ can be used to infer about the underlying average, across the three markers, of the differences between the difficulties of Section A questions whilst $\mathcal{M}(Y_{2t})$ and $\mathcal{M}(Y_{3t})$  looks at the differences between the markers of these Section A differences.  In Section \ref{sec:exvar}, $Tot_{g}$, the underlying total score for the $g$th marker, was noted to be highly correlated with $\mathcal{M}(W_{g7})$.  We now observe that $Tot = \frac{1}{3} \sum_{g=1}^{3} Tot_{g}$, the average across the three markers of the underlying total score, is highly correlated with $\mathcal{M}(Y_{17})$ (and only otherwise correlated with $\mathcal{M}(Y_{18})$) so that its' resolution given $\mathcal{C}(N)$ is $Res_{\mathcal{C}(N)}(Tot) = 0.9272\lambda_{17}+0.0728\lambda_{18}$. Similarly,  for $g \neq h$,  the difference in totals between marker $g$ and $h$, $Tot_{g} - Tot_{h}$,  has a resolution given by $Res_{\mathcal{C}(N)}(Tot_{g} - Tot_{h}) = 0.9272\lambda_{27}+0.0728\lambda_{28}$. 
\section{Sampling from finite populations \label{sec:finite}}

\subsection{Finite representation theorem\label{sec:4.1}}
We now consider the scenario when each group contains a finite, rather than an infinite, number of individuals. Suppose that the $g$th group contains a total of $m_{g}$ individuals and that the relationships between individuals are given by Assumptions \ref{assum1} and \ref{assum2}. Thus, the only modelling difference from Section \ref{sec:sampinf} is that individuals belong to finite rather than infinite populations. Consequently, the representation theorem given by (\ref{eq:as1}) no longer applies and instead we must utilise the finite representation theorem of Goldstein (1986)\nocite{Goldstein:1986}:
\begin{eqnarray}
\mathcal{C}_{gi} & = & \widetilde{\mathcal{M}}(\mathcal{C}_{g}) + \widetilde{\mathcal{R}}_{i}(\mathcal{C}_{g}) \label{eq:finrep}
\end{eqnarray}
where $\widetilde{\mathcal{M}}(\mathcal{C}_{g}) = \frac{1}{m_{g}} \sum_{i=1}^{m_{g}} \mathcal{C}_{gi}$ is the population mean and $\widetilde{\mathcal{R}}_{i}(\mathcal{C}_{g}) = \mathcal{C}_{gi} - \widetilde{\mathcal{M}}(\mathcal{C}_{g})$ the residual vector for the $i$th individual in the $g$th group. For all $g$, $h$, $i$, $j$, $\widetilde{\mathcal{R}}_{i}(\mathcal{C}_{g})$ is uncorrelated with $\widetilde{\mathcal{M}}(\mathcal{C}_{h})$ and, for $g \neq h$, uncorrelated with $\widetilde{\mathcal{R}}_{j}(\mathcal{C}_{h})$. The essential difference between (\ref{eq:as1}) and (\ref{eq:finrep}) is that the finite nature of the population induces a correlation between $\widetilde{\mathcal{R}}_{i}(\mathcal{C}_{g})$ and $\widetilde{\mathcal{R}}_{j}(\mathcal{C}_{g})$ for each $i \neq j$, the correlation being of order $\frac{1}{m_{g}}$. The consequence of the mean vectors remaining uncorrelated with the residuals is that the equivalent Bayes linear sufficiency conditions derived in the infinite case will hold in the finite setting and, as we shall show, leads to the similarities in approach between the two cases when adjusting the mean vectors. We collect the population sizes together into the matrix $M = diag(m_{1}, \ldots, m_{g_{0}})$ and the population mean vectors together as $\widetilde{\mathcal{M}}(\mathcal{C}) = \{\widetilde{\mathcal{M}}(\mathcal{C}_{1}), \ldots, \widetilde{\mathcal{M}}(\mathcal{C}_{g_{0}})\}$. We may thus deduce that whereas $Var(\mathcal{M}(\mathcal{C})) = A \otimes C$, $Var(\widetilde{\mathcal{M}}(\mathcal{C})) = (A \otimes C) + (M^{-1}B \otimes D) - (M^{-1}\hat{A} \otimes C)$.
\subsection{Adjustment of $\langle \widetilde{\mathcal{M}}(\mathcal{C}) \rangle$ given $\mathcal{C}(N)$ \label{sec:4.2}}
We now explore the impact of working with finite populations rather than infinite ones and so consider the effect of the observation of the sample $\mathcal{C}(N)$ on individual quantities in the population mean collection $\widetilde{\mathcal{M}}(\mathcal{C})$, comparing this with the corresponding adjustment for $\mathcal{M}(\mathcal{C})$ summarised by Corollary \ref{cor:full}. First we show that it is sufficient to work with the sample means, $\overline{\mathcal{C}}(N)$. Noting that, for all $g$, $h$, $j$, $Cov(\widetilde{\mathcal{M}}(\mathcal{C}_{g}), \mathcal{C}_{hj}) = Cov(\widetilde{\mathcal{M}}(\mathcal{C}_{g}), \widetilde{\mathcal{M}}(\mathcal{C}_{h})) = Cov(\widetilde{\mathcal{M}}(\mathcal{C}_{g}), \overline{\mathcal{C}}_{h})$ then, in an analogous fashion to Lemma \ref{lem:mean}, we have the following lemma.
\begin{Lem} \label{lem:fin1}
The collection $\overline{\mathcal{C}}(N)$ is Bayes linear sufficient for $\mathcal{C}(N)$ for adjusting $\widetilde{\mathcal{M}}(\mathcal{C})$. Thus, as $\langle \overline{\mathcal{C}}(N) \rangle \subseteq \langle \mathcal{C}(N) \rangle$, we have $E_{\mathcal{C}(N)}(\widetilde{\mathcal{M}}(\mathcal{C})) = E_{\overline{\mathcal{C}}(N)}(\widetilde{\mathcal{M}}(\mathcal{C}))$ and $Var_{\mathcal{C}(N)}(\widetilde{\mathcal{M}}(\mathcal{C})) = Var_{\overline{\mathcal{C}}(N)}(\widetilde{\mathcal{M}}(\mathcal{C}))$.
\end{Lem}
We shall show that, as in the finite case, this adjustment can be performed by considering separately the analysis of variables and of groups. The variable problem is identical to that in the infinite case and is one of sampling from a finite second-order exchangeable population. The group problem is analogous to that in the infinite case and is identical to it when the sampling fraction is the same in each group. In the balanced design, where each group contains $m$ individuals of which we sample $n$, the group problem can also be viewed as sampling from a finite second-order exchangeable population.
\subsubsection{Variable analysis: adjustment of $\langle \widetilde{\mathcal{M}}(\mathcal{C}_{g}) \rangle$ given $\mathcal{C}_{g}(n_{g})$ \label{sec:finvariables}}
In a corresponding fashion to Section \ref{sec:variables}, for each $g = 1, \ldots, g_{0}$ we explore the adjustment of quantities contained in $\langle \widetilde{\mathcal{M}}(\mathcal{C}_{g}) \rangle$ given $\mathcal{C}_{g}(n_{g})$. This is a problem of sampling $n_{g}$ individuals from a finite second-order exchangeable population of size $m_{g}$. For each $g$, create $\widetilde{\mathcal{M}}(W_{gt}) = \sum_{v=1}^{v_{0}} U_{vt}\widetilde{\mathcal{M}}(X_{gv})$ where $U_{vt}$ is the $v$th component of the $t$th canonical variable direction, as described in Definition \ref{def:var}. From Theorem 2 of Shaw and Goldstein (2012)\nocite{Shaw/Goldstein:2012} we have the finite version of Corollary \ref{cor:var}.
\begin{Cor} \label{cor:varfin}
For a sample of size $n_{g}$, $\mathcal{C}_{g}(n_{g})$, drawn from a population of size $m_{g}$, the collection $\widetilde{\mathcal{M}}(\mathcal{W}^{*}_{g}) = \{\widetilde{\mathcal{M}}(W_{g1}), \ldots, \widetilde{\mathcal{M}}(W_{gv_{0}})\}$ forms a basis for $\lincom{\widetilde{\mathcal{M}}(\mathcal{C}_{g})}$. The $\widetilde{\mathcal{M}}(W_{gs})$ are a priori uncorrelated, and, for all samples of any size, a posteriori uncorrelated and are the canonical directions for the adjustment. The posterior adjusted expectation, $\tilde{\mu}_{[g: n_{g}]s} = E_{\mathcal{C}_{g}(n_{g})}(\widetilde{\mathcal{M}}(W_{gt}))$, and posterior adjusted precision, $\tilde{r}_{[g: n_{g}]t} = Var^{-1}_{\mathcal{C}_{g}(n_{g})}(\widetilde{\mathcal{M}}(W_{gt}))$, are given by
\begin{eqnarray}
\tilde{\mu}_{[g: n_{g}]t} \ = \ \frac{\tilde{r}_{[g: 0]t}\mu_{[g:0]t}+a(m_{g}, n_{g})n_{g}\tilde{r}_{[g]t}\overline{W}_{gt}}{\tilde{r}_{[g: 0]t}+a(m_{g}, n_{g})n_{g}\tilde{r}_{[g]t}}; \ \tilde{r}_{[g: n_{g}]t} \ = \  \tilde{r}_{[g: 0]t}+a(m_{g}, n_{g})n_{g}\tilde{r}_{[g]t} \label{eq:varfin}
\end{eqnarray}
where $\tilde{r}_{[g: 0]t} = Var^{-1}(\widetilde{\mathcal{M}}(W_{gt}))$ and, for any $i$, $\tilde{r}_{[g]t} = Var^{-1}_{\widetilde{\mathcal{M}}(\mathcal{C}_{g})}(W_{gti})$ and $a(m_{g}, n_{g}) = \left(1 - \frac{n_{g}-1}{m_{g} - 1}\right)^{-1}$ is the finite population correction (fpc) for a sample of size $n_{g}$ drawn from a population of size $m_{g}$.
\end{Cor}
The $\widetilde{\mathcal{M}}(\mathcal{W}^{*}_{g})$ share the same coordinate representation as the $\mathcal{M}(\mathcal{W}^{*}_{g})$ of Corollary \ref{cor:var} and the univariate quantities in (\ref{eq:varfin}) are the finite equivalents of those in (\ref{eq:var}). Note how the fpc applies in the same way for each $t$. Observe both the computational simplicity and the ease of interpretation: the $\widetilde{\mathcal{M}}(\mathcal{W}^{*}_{g})$ have the same coordinate representation for each choice of $g$, $m_{g}$ and $n_{g}$ and are derived by solving the generalised eigenvalue problem given in Definition \ref{def:var} from which the quantitative information can also be directly obtained. If we denote by $\tilde{\lambda}_{[g:n_{g}]t}$ the resolution of $\widetilde{\mathcal{M}}(W_{gt})$ given $\mathcal{C}_{g}(n_{g})$ then
\begin{eqnarray}
\tilde{\lambda}_{[g:n_{g}]t} \ = \ \frac{n_{g}\{(m_{g}-1)\alpha_{gg} \phi_{t} + \gamma_{g}\}}{m_{g}\{(n_{g}-1)\alpha_{gg} \phi_{t} + \gamma_{g}\}} \ = \ \lambda_{[g:n_{g}]t} + \frac{n_{g}}{m_{g}}(1- \lambda_{[g:n_{g}]t}). \label{eq:resfin}
\end{eqnarray}
Thus, the increase in the resolution in the finite model occurs as an additive correction term to the corresponding resolution in the infinite model, the correction depending both upon the sampling fraction and the size of the (infinite) resolution, (\ref{eq:res}). Shaw and Goldstein (2012)\nocite{Shaw/Goldstein:2012} illustrate how the sampling fraction is also used to directly adjust the infinite model quantities given by (\ref{eq:var}) to obtain those of the finite model, (\ref{eq:varfin}). In particular, using (\ref{eq:resfin}), we may obtain, for each $t$ and $g$,
\begin{eqnarray}
\tilde{\mu}_{[g:n_{g}]t} \ = \ \left(1 - \frac{n_{g}}{m_{g}}\right)\mu_{[g:n_{g}]t} + \frac{n_{g}}{m_{g}} \overline{W}_{gt}; \ \frac{\tilde{r}_{[g: 0]t}}{\tilde{r}_{[g: n_{g}]t}} \ = \ \left(1 - \frac{n_{g}}{m_{g}}\right)\frac{r_{[g: 0]t}}{r_{[g: n_{g}]t}}. \label{eq:fincor}
\end{eqnarray}
We can also use (\ref{eq:resfin}) to compare variance reductions in differing groups. We have
\begin{eqnarray*}
\tilde{\lambda}_{[g:n_{g}]t} > \tilde{\lambda}_{[h:n_{h}]t} \ \Leftrightarrow \ \left(1 - \frac{n_{g}}{m_{g}}\right)\lambda_{[g:n_{g}]t}-\left(1 - \frac{n_{h}}{m_{h}}\right)\lambda_{[h:n_{h}]t}+\left(\frac{n_{g}}{m_{g}}-\frac{n_{h}}{m_{h}}\right) > 0. 
\end{eqnarray*}
If the sampling fraction is the same in each group, so that $N = \theta M$ for some $\theta \leq 1$, then $\tilde{\lambda}_{[g:n_{g}]t} - \tilde{\lambda}_{[h:n_{h}]t} = \left(1 - \theta\right)\left(\lambda_{[g:n_{g}]t}-\lambda_{[h:n_{h}]t}\right)$ so that the groups we learn most in agrees with the infinite model.
\subsubsection{Group analysis: adjustment of $\langle \widetilde{\mathcal{M}}(\mathcal{W}_{t})\rangle$ given $\overline{\mathcal{W}}_{t}$ \label{sec:fingroups}}
We mirror the approach of Section \ref{sec:groups} and consider adjustment across different groups. For each $t = 1, \ldots, v_{0}$ we form the collection $\widetilde{\mathcal{M}}(\mathcal{W}_{t}) = \{\widetilde{\mathcal{M}}(W_{1t}), \ldots, \widetilde{\mathcal{M}}(W_{g_{0}t})\}$ and form the total collection $\widetilde{\mathcal{M}}(\mathcal{W}) = \{\widetilde{\mathcal{M}}(\mathcal{W}_{1}), \ldots, \widetilde{\mathcal{M}}(\mathcal{W}_{v_{0}})\}$.
For each $t$ we consider the adjustment of $\langle \widetilde{\mathcal{M}}(\mathcal{W}_{t}) \rangle$ given $\overline{\mathcal{W}}_{t}$ and, in an identical fashion to Section \ref{sec:combined}, in Section \ref{sec:fincombined} show that this will enable us to completely determine the adjustment of $\langle \widetilde{\mathcal{M}}(\mathcal{C})\rangle$ given $\mathcal{C}(N)$. However, unlike in the variable analysis, as well as a quantitative difference there is a qualitative difference between the infinite and finite models: the canonical group directions obtained in Definition \ref{def:group} are, in general, not the canonical group directions for the finite problem. 

In Lemma \ref{lem:fin1} we showed that $\overline{\mathcal{C}}(N)$ is Bayes linear sufficient for $\mathcal{C}(N)$ for adjusting $\widetilde{\mathcal{M}}(\mathcal{C})$. In parallel to Lemma \ref{lem:meansuff}, the following lemma shows that for the adjustment of $\langle \widetilde{\mathcal{M}}(\mathcal{W}_{t})\rangle$ we may restrict attention to $\overline{\mathcal{W}}_{t}$ rather than $\overline{\mathcal{C}}(N)$. The result follows in a similar way to Lemma \ref{lem:fin1} by noting that, for each $t$, we have $Cov(\widetilde{\mathcal{M}}(\mathcal{W}_{t}), \overline{\mathcal{W}}_{t}) = Var(\widetilde{\mathcal{M}}(\mathcal{W}_{t})) = A + \phi_{t}^{-1}M^{-1}B -M^{-1}\hat{A}$ and $Cov(\widetilde{\mathcal{M}}(\mathcal{W}_{t}), \overline{\mathcal{C}}(N)) = \{A + \phi_{t}^{-1}M^{-1}B-M^{-1}\hat{A}\} \otimes U_{t}^{T}C$. 
\begin{Lem} \label{lem:meansufffin}
For each $t=1, \ldots, v_{0}$, the collection $\overline{\mathcal{W}}_{t}$ is Bayes linear sufficient for $\overline{\mathcal{C}}(N)$ for adjusting $\widetilde{\mathcal{M}}(\mathcal{W}_{t})$ so that, as $\langle \overline{\mathcal{W}}_{t} \rangle \subseteq \langle \overline{\mathcal{C}}(N) \rangle$, $E_{\overline{\mathcal{C}}(N)}(\widetilde{\mathcal{M}}(\mathcal{W}_{t})) = E_{\overline{\mathcal{W}}_{t}}(\widetilde{\mathcal{M}}(\mathcal{W}_{t}))$ and $Var_{\overline{\mathcal{C}}(N)}(\widetilde{\mathcal{M}}(\mathcal{W}_{t})) = Var_{\overline{\mathcal{W}}_{t}}(\widetilde{\mathcal{M}}(\mathcal{W}_{t}))$.
\end{Lem}
The canonical group directions and resolutions defined in Definition \ref{def:group} do not, in general, provide the wherewithal for a canonical analysis of the adjustment of $\langle \widetilde{\mathcal{M}}(\mathcal{W}_{t}) \rangle$ given $\overline{\mathcal{W}}_{t}$. In the finite setting we require the following definition.
\begin{Def} \label{fincangroup}
For each $t = 1, \ldots, v_{0}$ the $(s, t)$th finite canonical group direction is defined to be the $s$th column of the matrix $\tilde{V}_{t} = [\tilde{V}_{1t} \ldots \tilde{V}_{g_{0}t}]$ solving the generalised eigenvalue problem
\begin{eqnarray}
\{A + \phi_{t}^{-1}M^{-1}B - M^{-1}\hat{A}\}\tilde{V}_{t} = \{A + \phi_{t}^{-1}N^{-1}B - N^{-1}\hat{A}\}\tilde{V}_{t}\tilde{\Lambda}_{t} \nonumber
\end{eqnarray}
where $\phi_{t}$ is the $t$th underlying canonical variable resolution and $\tilde{\Lambda}_{t} = diag(\tilde{\lambda}_{1t}, \ldots, \tilde{\lambda}_{g_{0}t})$ is the matrix of eigenvalues, ordered so that $1 > \tilde{\lambda}_{1t} \geq \cdots \geq \tilde{\lambda}_{g_{0}t} > 0$. $\tilde{\lambda}_{st}$ is termed the $(s, t)$th finite canonical group resolution. $\tilde{V}_{t}$ is normed so that $\tilde{V}_{t}^{T}\{A + \phi_{t}^{-1}M^{-1}B- M^{-1}\hat{A}\}\tilde{V}_{t}$ $= I$, $\tilde{V}_{t}^{T}\{A + \phi_{t}^{-1}N^{-1}B - N^{-1}\hat{A}\}\tilde{V}_{t}\tilde{\Lambda}_{t} = I$. 
\end{Def} 
Let $\tilde{V}_{gst}$ denote the $g$th component of the $(s, t)$th finite canonical group direction and, for each $s=1, \ldots, g_{0}$, create $\widetilde{\mathcal{M}}(Z_{st}) = \sum_{g=1}^{g_{0}} \tilde{V}_{gst}\widetilde{\mathcal{M}}(W_{gt})$ and $\overline{Z}_{st}  = \sum_{g=1}^{g_{0}} \tilde{V}_{gst}\overline{W}_{gt}$. The following theorem is the finite equivalent to Theorem \ref{theo:group}; the proof is in the appendix.
\begin{Theo} \label{theo:groupfin} 
For each $t=1, \ldots, v_{0}$ the collection $\widetilde{\mathcal{M}}(\mathcal{Z}_{t}) = \{\widetilde{\mathcal{M}}(Z_{1t}), \ldots, \widetilde{\mathcal{M}}(Z_{g_{0}t})\}$ forms a basis for $\langle \widetilde{\mathcal{M}}(\mathcal{W}_{t}) \rangle$. The $\widetilde{\mathcal{M}}(Z_{st})$ are a priori uncorrelated, and, given $\overline{\mathcal{W}}_{t}$, a posteriori uncorrelated and are the canonical directions for the adjustment of $\langle \widetilde{\mathcal{M}}(\mathcal{W}_{t}) \rangle$. The posterior adjusted expectation, $\tilde{\mu}_{Nst} = E_{\overline{\mathcal{W}}_{t}}(\widetilde{\mathcal{M}}(Z_{st}))$, and posterior adjusted precision, $\tilde{r}_{Nst} = Var^{-1}_{\overline{\mathcal{W}}_{t}}(\widetilde{\mathcal{M}}(Z_{st}))$, are given by
\begin{eqnarray}
\tilde{\mu}_{Nst} \ = \ \frac{\tilde{r}_{0st}\tilde{\mu}_{0st}+ \tilde{r}_{st}\overline{Z}_{st}}{\tilde{r}_{0st} + \tilde{r}_{st}}; \ \tilde{r}_{Nst} \ = \ \tilde{r}_{0st} + \tilde{r}_{st}, \label{eq:groupfin}
\end{eqnarray}
where $\tilde{\mu}_{0st} = E(\widetilde{\mathcal{M}}(Z_{st}))$, $\tilde{r}_{0st} = Var^{-1}(\widetilde{\mathcal{M}}(Z_{st}))$ and $\tilde{r}_{st} = Var^{-1}_{\widetilde{\mathcal{M}}(\mathcal{C})}(\overline{Z}_{st})$. The resolution of $\widetilde{\mathcal{M}}(Z_{st})$ given $\overline{\mathcal{W}}_{t}$ is $\tilde{\lambda}_{st}$.
\end{Theo}
We observe that whilst the structural form of $\tilde{\mu}_{Nst}$ and $\tilde{r}_{Nst}$ in (\ref{eq:groupfin}) match those of $\mu_{Nst}$ and $r_{Nst}$ in (\ref{eq:group}), there is the crucial difference that the $\widetilde{\mathcal{M}}(Z_{st})$ do not, in general, share the same coordinate representation as the $\mathcal{M}(Y_{st})$. However, if the sampling fraction is the same in each group then the qualitative features of the adjustment of $\langle \widetilde{\mathcal{M}}(\mathcal{W}_{t})\rangle$ given $\overline{\mathcal{W}}_{t}$ are the same as those for the adjustment of $\langle \mathcal{M}(\mathcal{W}_{t})\rangle$ given $\overline{\mathcal{W}}_{t}$ as the following corollary demonstrates.
\begin{Cor} \label{cor:sampfrac}
If $N = \theta M$ for some $\theta \leq 1$ then, up to normalising constants, the $(s, t)$th finite canonical group direction, see Definition \ref{fincangroup}, is the same as the $(s, t)$th canonical group direction, see Definition \ref{def:group}. We have $\widetilde{\mathcal{M}}(Z_{st}) = \sqrt{\frac{\lambda_{st}}{\lambda_{st} + \theta (1-\lambda_{st})}} \widetilde{\mathcal{M}}(Y_{st})$ where $\widetilde{\mathcal{M}}(Y_{st})$ shares the same coordinate representation as $\mathcal{M}(Y_{st})$ and
\begin{eqnarray}
E_{\overline{\mathcal{W}}_{t}}(\widetilde{\mathcal{M}}(Y_{st})) \ = \ (1 - \theta)\mu_{Nst} + \theta \overline{Y}_{st}; \ Res_{\overline{\mathcal{W}}_{t}}(\widetilde{\mathcal{M}}(Y_{st})) \ = \ \tilde{\lambda}_{st} \ = \ \lambda_{st} + \theta(1-\lambda_{st}). \label{eq:sampfrac}
\end{eqnarray}
\end{Cor}
We thus see the role of $\theta$ in determining whether or not the finite nature of the populations is ignorable. The quantities in (\ref{eq:sampfrac}) share the same form as those in (\ref{eq:resfin}) and (\ref{eq:fincor}) so that $\theta$ replicates the role of the sampling fraction in the single group analysis when we were considering sampling second-order exchangeable populations. For all $s$ and $t$ the adjusted expectation of $\widetilde{\mathcal{M}}(Y_{st})$ is a weighted average of the adjusted expectation of $\mathcal{M}(Y_{st})$ and the observed mean $\overline{Y}_{st}$, the weights being dependent upon the ratio of the total population observed in the sample. The resolution of each $\widetilde{\mathcal{M}}(Y_{st})$ has an additive correction term to that of the resolution for $\mathcal{M}(Y_{st})$: the size of the correction depending both on the sampling fraction $\theta$ and the initial size of the resolution. A smaller correction term is applied to those quantities $\widetilde{\mathcal{M}}(Y_{st})$ for which the corresponding quantities $\mathcal{M}(Y_{st})$ have the largest resolutions, that is those that we learn most about.

Recall that, from the motivation of Corollary \ref{cor:balanced}, in the balanced design the adjustment of each $\langle \mathcal{M}(\mathcal{W}_{t}) \rangle$ given $\overline{\mathcal{W}}_{t}$ can be interpreted as one of sampling $n$ individuals from an infinitely exchangeable second-order population. In the finite case, when $m_{g} = m$ and $n_{g} = n$ for all $g = 1, \ldots, g_{0}$, the adjustment of each $\langle \widetilde{\mathcal{M}}(\mathcal{W}_{t}) \rangle$ given $\overline{\mathcal{W}}_{t}$ is one of sampling $n$ individuals from a finitely exchangeable second-order population of size $m$. To see this, for each $t = 1, \ldots, v_{0}$ and each $i = 1, \ldots, m$ form the collection $\mathcal{W}_{ti} = \{W_{1ti}, \ldots, W_{g_{0}ti}\}$. The $\mathcal{W}_{t1}, \ldots, \mathcal{W}_{tm}$ are second-order exchangeable over the individuals so, using the representation theorem of Goldstein (1986)\nocite{Goldstein:1986}, we can write $\mathcal{W}_{ti} = \widetilde{\mathcal{M}}(\mathcal{W}_{t}) + \widetilde{\mathcal{R}}_{i}(\mathcal{W}_{t})$. If we observe a sample of size $n$, $\mathcal{W}_{t1}, \ldots, \mathcal{W}_{tn}$, then the sample mean $\overline{\mathcal{W}}_{t}$ is Bayes linear sufficient for the adjustment of $\langle \widetilde{\mathcal{M}}(\mathcal{W}_{t}) \rangle$. Using Theorem 2 of Shaw and Goldstein (2012)\nocite{Shaw/Goldstein:2012} we have the the finite version of Corollary \ref{cor:balanced}.  
\begin{Cor} \label{cor:balancedfin} If $N= nI$ and $M = mI$ then, up to normalising constants, the $(s, t)$th finite canonical group direction is the same as the $(s, t)$th canonical group direction and can be obtained as the $s$th column of the matrix $V_{t}$ solving $AV_{t} = (A + \phi_{t}^{-1}B - \hat{A})V_{t}\Psi_{t}$. The posterior adjusted expectation $\tilde{\mu}_{Nst}$, posterior adjusted precision $\tilde{r}_{Nst}$ and resolution $\tilde{\lambda}_{Nst}$ of each $\widetilde{\mathcal{M}}(Y_{st})$ given $\overline{\mathcal{W}}_{t}$ are given by
\begin{eqnarray*} 
& &\tilde{\mu}_{Nst} \ = \ \frac{\tilde{r}_{0st}\tilde{\mu}_{0st} + a(m, n)n\tilde{q}_{st}\overline{Y}_{st}}{\tilde{r}_{0st}+ a(m, n)n\tilde{q}_{st}} \ = \ \left(1 - \frac{n}{m}\right)\mu_{0st} + \frac{n}{m}\overline{Y}_{st}; \\
& &\tilde{r}_{Nst} \ = \ \tilde{r}_{0st}+ a(m, n)n\tilde{r}_{(1)st}; \ \tilde{\lambda}_{st} \ = \  \lambda_{st} + \frac{n}{m}(1 - \lambda_{st})
\end{eqnarray*}
where $\tilde{q}_{st} = Var^{-1}_{\widetilde{\mathcal{M}}(\mathcal{C})}(Y_{sti})$, $Y_{sti} = \sum_{g=1}^{g_{0}} V_{gs}W_{gti}$ shares the same coordinate representation as $\widetilde{\mathcal{M}}(Y_{st})$, $\mu_{0st}$ and $\lambda_{st}$ are as given in Corollary \ref{cor:balanced} and $a(m, n) = \left(1 - \frac{n-1}{m-1}\right)^{-1}$ is the fpc.
\end{Cor}
In contrast to the infinite case described in Corollary \ref{cor:prop}, if $\alpha_{gg}/\gamma_{g} = a$ then the finite canonical group directions will typically still depend upon $t$. However, if additionally $N = \theta M$ then we can extend Corollary \ref{cor:sampfrac} to obtain the finite equivalent of Corollary \ref{cor:prop}: the $\widetilde{\mathcal{M}}(Y_{st})$ share, up to normalising constants, the same coordinate representation for each $t$ derived as the $s$th column of the matrix $V$ solving $AV = (A + N^{-1}B)V\Psi$ with $\tilde{\lambda}_{st}$ as in (\ref{eq:sampfrac}) and $\lambda_{st}$ as in (\ref{eq:separable}). If we have the balanced design described in Corollary \ref{cor:balancedfin} then, for all choices of $n$, $m$ and $t$, $\widetilde{\mathcal{M}}(Y_{st})$ and $\tilde{\lambda}_{st}$ can be obtained from the same $g_{0} \times g_{0}$ problem $AV = (A + B)V\Psi_{(1)}$ as $\mathcal{M}(Y_{st})$ and $\lambda_{st}$, with $\lambda_{st}$ given by (\ref{eq:sepbalanced}). These results correspond to those which would be obtained using the covariance representations over individuals induced by the model proposed in Shaw and Goldstein (1999)\nocite{Shaw/Goldstein:1999} as described in Section \ref{sec:3.1.2}.

\subsubsection{Variable and group analysis combined: adjustment of $\langle \widetilde{\mathcal{M}}(\mathcal{C})\rangle$ given $\mathcal{C}(N)$ \label{sec:fincombined}}
As with the infinite model, we retain the elegance that, having first solved the canonical variable problem, solving each of the canonical group problems will enable us to completely determine the adjustment of $\langle \widetilde{\mathcal{M}}(\mathcal{C}) \rangle$ given $\mathcal{C}(N)$. To show this we utilise the following lemma whose proof is in the appendix. 
\begin{Lem} \label{lem:sepfin}
1. The collection $\widetilde{\mathcal{M}}(\mathcal{W}) = \{\widetilde{\mathcal{M}}(\mathcal{W}_{1}), \ldots, \widetilde{\mathcal{M}}(\mathcal{W}_{v_{0}})\}$ is Bayes linear sufficient for $\overline{\mathcal{C}}(N)$ for adjusting $\widetilde{\mathcal{M}}(\mathcal{C})$ so that
\begin{eqnarray*}
E_{\overline{\mathcal{C}}(N)}(\widetilde{\mathcal{M}}(\mathcal{C})) & = & E_{\overline{\mathcal{C}}(N)}\{E_{\widetilde{\mathcal{M}}(\mathcal{W})}(\widetilde{\mathcal{M}}(\mathcal{C}))\}; \\
Var_{\overline{\mathcal{C}}(N)}(\widetilde{\mathcal{M}}(\mathcal{C})) & = & Var_{\widetilde{\mathcal{M}}(\mathcal{W})}(\widetilde{\mathcal{M}}(\mathcal{C})) +  Var_{\overline{\mathcal{C}}(N)}\{E_{\widetilde{\mathcal{M}}(\mathcal{W})}(\widetilde{\mathcal{M}}(\mathcal{C}))\}.
\end{eqnarray*}

2. For all $t \neq t'$, the collection $\overline{\mathcal{C}}(N)$ is Bayes linear sufficient for $\widetilde{\mathcal{M}}(\mathcal{W}_{t'})$ for adjusting $\widetilde{\mathcal{M}}(\mathcal{W}_{t})$ so that $Cov_{\overline{\mathcal{C}}(N)}(\widetilde{\mathcal{M}}(\mathcal{W}_{t}), \widetilde{\mathcal{M}}(\mathcal{W}_{t'})) = 0$.
\end{Lem}
Mirroring the discussion to Lemma \ref{lem:sep}, the first part of Lemma \ref{lem:sepfin} demonstrates that the adjustment of $\langle \widetilde{\mathcal{M}}(\mathcal{W}) \rangle$ given $\overline{\mathcal{C}}(N)$ is sufficient to obtain the adjustment of $\langle \widetilde{\mathcal{M}}(\mathcal{C}) \rangle$ whilst the second part of Lemma \ref{lem:sepfin}, combined with Lemma \ref{lem:meansufffin}, shows that $Var_{\overline{\mathcal{C}}(N)}(\widetilde{\mathcal{M}}(\mathcal{W})) = \oplus_{t=1}^{v_{0}} Var_{\overline{\mathcal{W}}_{t}}(\widetilde{\mathcal{M}}(\mathcal{W}_{t}))$ and $T_{\widetilde{\mathcal{M}}(\mathcal{W}) \, : \, \overline{\mathcal{C}}(N)} = \oplus_{t=1}^{v_{0}} T_{\widetilde{\mathcal{M}}(\mathcal{W}_{t}) \, : \, \overline{\mathcal{W}}_{t}}$.
 Putting this together with Lemma \ref{lem:fin1} gives the following corollary to Theorem \ref{theo:groupfin}. 
\begin{Cor} \label{cor:fullfin}
The collection $\widetilde{\mathcal{M}}(\tilde{\mathcal{Z}}) = \{\widetilde{\mathcal{M}}(Z_{st}), s=1, \ldots, g_{0}, t = 1, \ldots, v_{0}\}$ forms a basis for $\langle \widetilde{\mathcal{M}}(\mathcal{C}) \rangle$. The $\widetilde{\mathcal{M}}(Z_{st})$ are a priori uncorrelated, and, given $\mathcal{C}(N)$, a posteriori uncorrelated and are the canonical directions for the adjustment of $\langle \widetilde{\mathcal{M}}(\mathcal{C}) \rangle$.
\end{Cor}
Corollary \ref{cor:fullfin} is the finite equivalent to Corollary \ref{cor:full} and we note the similarities and differences between the two when considering the adjustment of the mean structure given $\mathcal{C}(N)$. In each case, we exploit the (generalised) conditional independence structure to separate the problem into two parts: firstly that of solving a canonical variable problem and secondly then considering the adjustment across all of the groups for each canonical variable direction. Thus, in both the finite and infinite cases, we separate the $g_{0}v_{0} \times g_{0}v_{0}$ problem into a single $v_{0} \times v_{0}$ problem, which does not depend upon either the sample size or the population size, and $v_{0}$ $g_{0} \times g_{0}$ group problems which do depend upon the sample and population size.

\subsection{Example illustration\label{ex:fin}}
We return to the examination data example discussed in Sections \ref{sec:ex1}, \ref{sec:exvar} and \ref{sec:exinf}. The examination is sat by 355 candidates and the first marker can mark for one day, the second for two days, and the third for four days. With this in mind, the examiner allocated $m_{1} = 51$ scripts to the first marker, $m_{2} = 101$ to the second and $m_{3} = 203$ to the third. Consequently, the population sizes he samples from are finite and thus he performs his analysis using the results of Sections \ref{sec:finvariables}-\ref{sec:fincombined}.

He first performs the variable analysis. From Corollary \ref{cor:varfin} he just modifies the analysis of Section \ref{sec:exvar}, working with the variables $\widetilde{\mathcal{M}}(\mathcal{C}_{g})$ rather that $\mathcal{M}(\mathcal{C}_{g})$. Hence, for example,
\begin{eqnarray}
\widetilde{\mathcal{M}}(W_{gt}) & \propto & \widetilde{\mathcal{M}}(X_{gt+1}) - \frac{1}{t}\sum_{v=1}^{t} \widetilde{\mathcal{M}}(X_{gv}), \ t = 1, \ldots, 4 \label{eq:secAdifffin}
\end{eqnarray}
is the finite equivalent to the infinite quantities given in equation (\ref{eq:secAdiff}) and so represent differences between the difficulties of Section A questions under marker $g$. For a sample of $n_{g}$ in the $g$th group we find, using (\ref{eq:resfin}),
\begin{eqnarray*}
\tilde{\lambda}_{[g:n_{g}]t} \ = \ \lambda_{[g:n_{g}]t} + \frac{n_{g}}{m_{g}}(1- \lambda_{[g:n_{g}]t}) \ = \ \frac{n_{g}\phi_{t}}{(n_{g}-1)\phi_{t} +1} + \frac{n_{g}}{m_{g}}\left(\frac{1-\phi_{t}}{(n_{g}-1)\phi_{t} +1}\right)
\end{eqnarray*}
where the $\phi_{t}$s are given in (\ref{eq:phis}).

The examiner now turns to the group analysis and, as in Section \ref{sec:exinf}, considers taking a sample of $n$ scripts from each marker and forms, for each $t = 1, \ldots, 8$, the collections $\widetilde{\mathcal{M}}(\mathcal{W}_{t}) = \{\widetilde{\mathcal{M}}(W_{1t}), \ldots, \widetilde{\mathcal{M}}(W_{3t})\}$ and $\overline{\mathcal{W}}_{t} = \{\overline{W}_{1t}, \ldots, \overline{W}_{3t}\}$. In full generality, the examiner would have to solve the eight $3 \times 3$ generalised eigenvalue problems given in Definition \ref{fincangroup}. Despite the balanced sample, the unbalanced finite sample sizes, as $M = diag(51, 101, 203)$, mean that Corollary \ref{cor:sampfrac} does not apply: the $(s, t)$th finite canonical group direction will differ from the corresponding canonical group direction. However, as there are only four distinct $\phi_{t}$s there will only be four different problems to solve. Even with the simplified choice of $A$ and $B$, the explicit derivation of the eigenstructure is not straightforward. However, for this example, we can still gain insight into the relationship between the finite and infinite if we consider the resolved uncertainty for $\langle \widetilde{\mathcal{M}}(\mathcal{W}_{t}) \rangle$ given $\overline{\mathcal{W}}_{t}$, denoted $RU_{\overline{\mathcal{W}}_{t}}(\widetilde{\mathcal{M}}(\mathcal{W}_{t}))$. This is, see Definition 3.19 of Goldstein and Wooff (2007)\nocite{Goldstein/Wooff:2007}, the sum of the $g_{0} =3$ $t$th finite canonical group resolutions. From Theorem \ref{theo:groupfin} we have, with $N = diag(n, n, n)$, 
\begin{eqnarray}
RU_{\overline{\mathcal{W}}_{t}}(\widetilde{\mathcal{M}}(\mathcal{W}_{t})) & = & \sum_{s=1}^{g_{0}} \tilde{\lambda}_{st} \ = \ \mbox{trace}(T_{\tilde{\mathcal{M}}(\mathcal{W}_{t}) \, : \, \overline{\mathcal{W}}_{t}}) \nonumber \\
& = & \mbox{trace}\left[\{A + \phi_{t}^{-1}N^{-1}B -N^{-1}\hat{A}\}^{-1}\{A + \phi_{t}^{-1}M^{-1}B - M^{-1}\hat{A}\}\right] \nonumber \\
& = & \sum_{s=1}^{g_{0}} \lambda_{st} + \left(\frac{1}{g_{0}}\sum_{g=1}^{g_{0}} \frac{n}{m_{g}}\right) \sum_{s=1}^{g_{0}} (1-\lambda_{st}). \label{eq:exam6}
\end{eqnarray}
Thus, for the balanced design, with the simplified choice of $A$ and $B$ reflecting group exchangeability, we can see the generalisation, compare (\ref{eq:exam6}) to (\ref{eq:sampfrac}), of the role of the sampling fractions $n/m_{g}$ in the finite model: the resolved uncertainty for $\langle \widetilde{\mathcal{M}}(\mathcal{W}_{t}) \rangle$ given $\overline{\mathcal{W}}_{t}$ is the resolved uncertainty for $\langle \mathcal{M}(\mathcal{W}_{t}) \rangle$ given $\overline{\mathcal{W}}_{t}$ plus a correction term depending both upon the average sampling fraction and the total remaining uncertainty for $\langle \mathcal{M}(\mathcal{W}_{t}) \rangle$ given $\overline{\mathcal{W}}_{t}$. Note that this is the form for each $t = 1, \ldots 8$. 

For illustrative purposes, we consider the cases for $\phi_{1} = \cdots = \phi_{4} = 0.8$ and $\phi_{7} = 0.1666$ when $n = 10$. For $t = 1, \ldots, 4$, $\widetilde{\mathcal{M}}(W_{gt})$ is as given in (\ref{eq:secAdifffin}) and the examiner obtains:
\begin{eqnarray*}
\tilde{\lambda}_{1t} = 0.9919 & \mbox{with} & \widetilde{\mathcal{M}}(Z_{1t}) = 0.3871\widetilde{\mathcal{M}}(W_{1t}) + 0.3426\widetilde{\mathcal{M}}(W_{2t})+0.3236\widetilde{\mathcal{M}}(W_{3t}), \\
\tilde{\lambda}_{2t} = 0.8797 & \mbox{with} & \widetilde{\mathcal{M}}(Z_{2t}) = 2.0439\widetilde{\mathcal{M}}(W_{1t}) -1.3387\widetilde{\mathcal{M}}(W_{2t})-0.7243\widetilde{\mathcal{M}}(W_{3t}), \\
\tilde{\lambda}_{3t} = 0.8674 & \mbox{with} & \widetilde{\mathcal{M}}(Z_{3t}) = -0.3537\widetilde{\mathcal{M}}(W_{1t}) -1.6066\widetilde{\mathcal{M}}(W_{2t})+1.9701\widetilde{\mathcal{M}}(W_{3t}).
\end{eqnarray*}
For $t = 7$, $\widetilde{\mathcal{M}}(W_{g7}) = 0.2683TotalA_{g} + 0.0702TotalB_{g}$ where $TotalA_{g} = \sum_{v=1}^{5} \widetilde{\mathcal{M}}(X_{gv})$, $TotalB_{g} = \sum_{v=6}^{8} \widetilde{\mathcal{M}}(X_{gv})$ denote, respectively, the underlying script total on Section A and Section B for the $g$th marker. As in Section \ref{sec:exinf}, this quantity is heavily correlated with the underlying script total for the $g$th marker.  The examiner finds that:
\begin{eqnarray*}
\tilde{\lambda}_{17} = 0.8625 & \mbox{with} & \widetilde{\mathcal{M}}(Z_{17}) = 0.3894\widetilde{\mathcal{M}}(W_{17}) + 0.3370\widetilde{\mathcal{M}}(W_{27})+0.3153\widetilde{\mathcal{M}}(W_{37}), \\
\tilde{\lambda}_{27} = 0.3515 & \mbox{with} & \widetilde{\mathcal{M}}(Z_{27}) = 1.6610\widetilde{\mathcal{M}}(W_{17}) -1.1180\widetilde{\mathcal{M}}(W_{27})-0.6019\widetilde{\mathcal{M}}(W_{37}), \\
\tilde{\lambda}_{37} = 0.2856 & \mbox{with} & \widetilde{\mathcal{M}}(Z_{37}) = -0.3186\widetilde{\mathcal{M}}(W_{17}) -1.4398\widetilde{\mathcal{M}}(W_{27})+1.7911\widetilde{\mathcal{M}}(W_{37}).
\end{eqnarray*}
We see that whilst the directions do not share the same co-ordinate representation as those for the infinite model they are closely related. For example, using the co-ordinate representation of the $(1, t)$th canonical group direction gives $\widetilde{\mathcal{M}}(Y_{1t}) \propto \widetilde{\mathcal{M}}(W_{1t}) + \widetilde{\mathcal{M}}(W_{2t}) + \widetilde{\mathcal{M}}(W_{3t})$, the total (or average) across the markers of the $\widetilde{\mathcal{M}}(W_{gt})$. For $t = 1, \ldots, 4$, we have $\widetilde{\mathcal{M}}(Y_{1t}) \propto 2.8471\widetilde{\mathcal{M}}(Z_{1t})-0.0460\widetilde{\mathcal{M}}(Z_{2t})+0.02231\widetilde{\mathcal{M}}(Z_{3t})$ whilst, for $t = 7$,  $\widetilde{\mathcal{M}}(Y_{17}) \propto 2.8752\widetilde{\mathcal{M}}(Z_{17})-0.0662\widetilde{\mathcal{M}}(Z_{27})+0.0299\widetilde{\mathcal{M}}(Z_{37})$. In each case, we observe that $\widetilde{\mathcal{M}}(Y_{1t})$ is strongly correlated with $\widetilde{\mathcal{M}}(Y_{1t})$ and thus, in the finite model, we are retaining the insight that we expect to learn most about quantities corresponding to the total (or average) across the markers. Similarly, we observe that the coefficients of both the $\widetilde{\mathcal{M}}(Z_{2t})$ and $\widetilde{\mathcal{M}}(Z_{3t})$ sum almost to zero, matching the property of the corresponding $\widetilde{\mathcal{M}}(Y_{2t})$ and $\widetilde{\mathcal{M}}(Y_{3t})$.  Notice that, unlike in the infinite case, we have three distinct canonical group resolutions and that the difference between $\tilde{\lambda}_{2t}$ and $\tilde{\lambda}_{3t}$ seems to increase as $\phi_{t}$ decreases (for $\phi_{8} = 0.1133$ we find that $\tilde{\lambda}_{18} = 0.8024$, $\tilde{\lambda}_{28} = 0.2923$, and $\tilde{\lambda}_{38} = 0.2207$). We can however, utilise the resolution partition, see (\ref{eq:respart}), so that for any $Z \in \langle \widetilde{\mathcal{M}}(\mathcal{W}_{t}) \rangle$, $\tilde{\lambda}_{3t} \leq Res_{\overline{\mathcal{W}}_{t}}(Z)  \leq  \tilde{\lambda}_{1t}$ and, using Corollary \ref{cor:fullfin} as $Z \in \langle \widetilde{\mathcal{M}}(\mathcal{W}_{t}) \rangle$, $\tilde{\lambda}_{3t} \leq Res_{\mathcal{C}(N)}(Z)  \leq  \tilde{\lambda}_{1t}$.

This analysis is reassuring as it suggests that the canonical group directions obtained from Definition \ref{def:group} act as a close approximation to those obtained from Definition \ref{fincangroup} in this case where the sampling fractions are not too high (in this case, the largest sampling fraction is $10/51$). Notice that as $n$ increases towards $m_{1} = 51$ then the finite reality means that we will increasingly learn more and more about  the $\widetilde{\mathcal{M}}(W_{1t})$ as we observe an ever increasing fraction of the scripts of the first marker. For example, for $n=20$ we have $\widetilde{\mathcal{M}}(Z_{17}) = 0.4411\widetilde{\mathcal{M}}(W_{17}) + 0.3187\widetilde{\mathcal{M}}(W_{27})+0.2792\widetilde{\mathcal{M}}(W_{37})$, whilst for $n=40$ we have $\widetilde{\mathcal{M}}(Z_{17}) = 0.6574\widetilde{\mathcal{M}}(W_{17}) + 0.2070\widetilde{\mathcal{M}}(W_{27})+0.1532\widetilde{\mathcal{M}}(W_{37})$ and for $n=50$, $\widetilde{\mathcal{M}}(Z_{17}) = 0.9177\widetilde{\mathcal{M}}(W_{17}) + 0.0281\widetilde{\mathcal{M}}(W_{27})+0.0188\widetilde{\mathcal{M}}(W_{37})$.

  \section{Concluding remarks}
One possible interpretation of the relationship between the modelling of infinite populations and, the equivalently defined, finite populations is that the latter could be viewed as a perturbation of the former, the size of the perturbation being related to the sampling fraction. As Shaw and Goldstein (2012)\nocite{Shaw/Goldstein:2012} show (see Corollaries \ref{cor:var} and \ref{cor:varfin} and equation (\ref{eq:resfin}) of this paper), in the case when the models are second-order exchangeable, the perturbation only affects the canonical resolutions but not the canonical directions. The qualitative features remain the same whilst the quantitative features are simply modified using the sampling fraction: the canonical resolutions in the finite case are those in the infinite case with the addition of a factor which depends on the sampling fraction and the size of the resolution. In a more complicated model, where we may be sampling from different populations, Corollaries \ref{cor:sampfrac} and \ref{cor:balancedfin} illustrate that this behaviour is repeated if we sample the same fraction in each group. Note that, see the proof of Theorem \ref{theo:group}, $T_{\mathcal{M}(\mathcal{W}_{t}):\overline{\mathcal{W}}_{t}} = \{A + \phi_{t}^{-1}N^{-1}B - N^{-1}\hat{A}\}^{-1}A$ whereas, see the proof of Theorem \ref{theo:groupfin}, $T_{\widetilde{\mathcal{M}}(\mathcal{W}_{t}):\overline{\mathcal{W}}_{t}} = \{A + \phi_{t}^{-1}N^{-1}B- N^{-1}\hat{A}\}^{-1}\{A + \phi_{t}^{-1}M^{-1}B- M^{-1}\hat{A}\}$. The eigenstructure of $T_{\widetilde{\mathcal{M}}(\mathcal{W}_{t}):\overline{\mathcal{W}}_{t}}$, obtained in Definition \ref{fincangroup}, can be interpreted as a perturbation of the eigenstructure of $T_{\mathcal{M}(\mathcal{W}_{t}):\overline{\mathcal{W}}_{t}}$, obtained via Definition \ref{def:group}, the perturbation caused by the addition of the matrix $\phi_{t}^{-1}M^{-1}B- M^{-1}\hat{A}$. The size of the perturbation will depend upon both $M$ and $\phi_{t}$. We observe that the smaller the value of $\phi_{t}$, corresponding to smaller information being learnt in the variable problem, the larger the perturbation. We could consider a careful perturbation analysis of the eigenvalues and eigenvectors of $T_{\widetilde{\mathcal{M}}(\mathcal{W}_{t}):\overline{\mathcal{W}}_{t}}$ to fully explore the role of $\phi_{t}$ and the sampling fractions in the perturbation. A similar analysis could be performed with $T_{\mathcal{M}(\mathcal{W}_{t}):\overline{\mathcal{W}}_{t}} = \{A + \phi_{t}^{-1}N^{-1}B - N^{-1}\hat{A}\}^{-1}A$ when compared to the balanced design which, from Corollary \ref{cor:balanced}, fixes the group problems for all sample sizes.
\section*{Appendix}
\textbf{Proof of Lemma \ref{lem:mean} -} Note that for all $g$, $h$, $j$, $Cov(\overline{\mathcal{C}}_{g}, \mathcal{C}_{hj}) = Cov(\overline{\mathcal{C}}_{g}, \overline{\mathcal{C}}_{h})$. Thus, 
\begin{eqnarray}
Cov(\overline{\mathcal{C}}(N), \mathcal{C}(N)) & = & [1^{T}_{n_{1}} \otimes Var_{1}(\overline{\mathcal{C}}(N)) \ \ldots \ 1^{T}_{n_{g_{0}}} \otimes Var_{g_{0}}(\overline{\mathcal{C}}(N))] \label{eq:appen01}
\end{eqnarray}
where $1^{T}_{s}$ denotes the $1 \times s$ vector of ones and $Var_{g}(\overline{\mathcal{C}}(N)) = Cov(\overline{\mathcal{C}}(N), \overline{\mathcal{C}}_{g})$. Denoting by $I_{s}$ the $s \times s$ identity matrix and by $I_{r, s}$ the $r$th column of $I_{s}$ then, using (\ref{eq:appen01}),
\begin{eqnarray*}
Var^{-1}(\overline{\mathcal{C}}(N))Cov(\overline{\mathcal{C}}(N), \mathcal{C}(N)) & = & [1^{T}_{n_{1}} \otimes I_{1, g_{0}} \otimes I_{v_{0}}  \ \ldots \ 1^{T}_{n_{g_{0}}} \otimes I_{g_{0}, g_{0}} \otimes I_{v_{0}}]. 
\end{eqnarray*}
Now, for all $g$, $h$, $j$, $Cov(\mathcal{M}(\mathcal{C}_{g}), \mathcal{C}_{hj}) = Cov(\mathcal{M}(\mathcal{C}_{g}), \mathcal{M}(\mathcal{C}_{h})) = Cov(\mathcal{M}(\mathcal{C}_{g}), \overline{\mathcal{C}}_{h})$ so that  
\begin{eqnarray*}
\lefteqn{Cov(\mathcal{M}(\mathcal{C}), \overline{\mathcal{C}}(N))Var^{-1}(\overline{\mathcal{C}}(N))Cov(\overline{\mathcal{C}}(N), \mathcal{C}(N)) \ =} \\
& & [1^{T}_{n_{1}} \otimes Cov(\mathcal{M}(\mathcal{C}), \mathcal{M}(\mathcal{C}_{1})) \ \ldots \ 1^{T}_{n_{g_{0}}} \otimes Cov(\mathcal{M}(\mathcal{C}), \mathcal{M}(\mathcal{C}_{g_{0}}))] \ = \ Cov(\mathcal{M}(\mathcal{C}), \mathcal{C}(N)).
\end{eqnarray*}
The result immediately follows from Theorem 5.20 of Goldstein and Wooff (2007).\nocite{Goldstein/Wooff:2007} \hfill $\Box$

\ \newline
\textbf{Proof of Lemma \ref{lem:meansuff} -} Noting that $\mathcal{M}(W_{gt}) = U_{t}^{T}\mathcal{M}(\mathcal{C}_{g})$ and, using (\ref{eq:as1}), $\overline{W}_{gt} = U_{t}^{T}\mathcal{M}(\mathcal{C}_{g}) + \frac{1}{n_{g}} \sum_{i=1}^{n_{g}} U_{t}^{T}\mathcal{R}_{i}(\mathcal{C}_{g}) = U_{t}^{T}\overline{\mathcal{C}}_{g}$ then, from the scalings in Definition \ref{def:var}, $Cov(\mathcal{M}(\mathcal{W}_{t}), \overline{\mathcal{W}}_{t}) = A$ and $Var(\overline{\mathcal{W}}_{t}) = A + \phi_{t}^{-1} N^{-1}B - N^{-1}\hat{A}$. Similarly, $Cov(\mathcal{M}(\mathcal{W}_{t}), \overline{\mathcal{C}}(N)) = A \otimes U_{t}^{T}C$ and $Cov(\overline{\mathcal{W}}_{t}, \overline{\mathcal{C}}(N)) = \{A + \phi_{t}^{-1} N^{-1}B - N^{-1}\hat{A}\} \otimes U_{t}^{T}C$. Direct multiplication then shows that $Cov(\mathcal{M}(\mathcal{W}_{t}), \overline{\mathcal{C}}(N)) = Cov(\mathcal{M}(\mathcal{W}_{t}), \overline{\mathcal{W}}_{t})Var^{-1}(\overline{\mathcal{W}}_{t})Cov(\overline{\mathcal{W}}_{t}, \overline{\mathcal{C}}(N))$ and the result follows from Theorem 5.20 of Goldstein and Wooff (2007).\nocite{Goldstein/Wooff:2007} \hfill $\Box$

\ \newline
\textbf{Proof of Theorem \ref{theo:group} -} From Assumption \ref{assum2} we deduce that $Cov(\mathcal{M}(\mathcal{C}), \overline{\mathcal{C}}(N)) = Var(\mathcal{M}(\mathcal{C}))$ $= A \otimes C$ and $Var(\overline{\mathcal{C}}(N)) =  (A \otimes C) + (N^{-1}B \otimes D) - (N^{-1}\hat{A} \otimes C)$. Hence, $Var_{\mathcal{M}(\mathcal{C})}(\overline{\mathcal{C}}(N)) = (N^{-1}B \otimes D) - (N^{-1}\hat{A} \otimes C)$. Noting that $\mathcal{M}(Y_{st}) = V_{s}^{T}\mathcal{M}(\mathcal{W}_{t}) = (V_{s}^{T} \otimes U_{t}^{T})\mathcal{M}(\mathcal{C})$ and $\overline{Y}_{st} = V_{s}^{T}\overline{\mathcal{W}}_{t} = (V_{s}^{T} \otimes U_{t}^{T})\overline{\mathcal{C}}(N)$ and, for any $i, j$, letting $\delta_{ij}$ denote the Kronecker delta, from Definitions \ref{def:var} and \ref{def:group}, for all $s, s', t, t'$ we have 
\begin{eqnarray}
Cov(\mathcal{M}(Y_{st}), \mathcal{M}(Y_{s't'})) & = & V_{s}^{T}AV_{s'} \otimes U_{t}^{T}CU_{t'} \ = \ \delta_{ss'}\delta_{tt'} \label{eq:appen1}\\
Cov_{\mathcal{M}(\mathcal{C})}(\overline{Y}_{st}, \overline{Y}_{s't'}) & = & V_{s}^{T}N^{-1}BV_{s'} \otimes U_{t}^{T}DU_{t'} - V_{s}^{T}N^{-1}\hat{A}V_{s'} \otimes U_{t}^{T}CU_{t'} \nonumber\\ & = & \delta_{ss'}\delta_{tt'} \frac{1 - \lambda_{st}}{\lambda_{st}}. \label{eq:appen2}
\end{eqnarray}
The resolution transform for the adjustment of $\langle \mathcal{M}(\mathcal{W}_{t}) \rangle$ given $\overline{\mathcal{W}}_{t}$ is, from (\ref{eq:adj3}), $T_{\mathcal{M}(\mathcal{W}_{t}):\overline{\mathcal{W}}_{t}} = Var^{-1}(\mathcal{M}(\mathcal{W}_{t}))Cov(\mathcal{M}(\mathcal{W}_{t}), \overline{\mathcal{W}}_{t})Var^{-1}(\overline{\mathcal{W}}_{t})Cov(\overline{\mathcal{W}}_{t}, \mathcal{M}(\mathcal{W}_{t}))$. By noting that $\mathcal{M}(\mathcal{W}_{t}) = (I_{g_{0}} \otimes U_{t}^{T})\mathcal{M}(\mathcal{C})$ we see that $Var(\mathcal{M}(\mathcal{W}_{t})) = A$ and thus, additionally using the beliefs stated in the proof of Lemma \ref{lem:meansuff}, we have
\begin{eqnarray}
T_{\mathcal{M}(\mathcal{W}_{t}):\overline{\mathcal{W}}_{t}} & = & \{A + \phi_{t}^{-1}N^{-1}B - N^{-1}\hat{A}\}^{-1}A \label{eq:appendix4}
\end{eqnarray}
which, from Definition \ref{def:group}, for each $s = 1, \ldots, g_{0}$, has eigenvector $V_{st}$ and corresponding eigenvalue $\lambda_{st}$. Now, see for example Property 3.17 of Goldstein and Wooff (2007)\nocite{Goldstein/Wooff:2007}, 
\begin{eqnarray}
Cov_{\overline{\mathcal{W}}_{t}}(\mathcal{M}(Y_{st}), \mathcal{M}(Y_{s't})) \ = \ V_{st}^{T}Var(\mathcal{M}(\mathcal{W}_{t}))\{I - T_{\mathcal{M}(\mathcal{W}_{t}):\overline{\mathcal{W}}_{t}}\}V_{s't} \ = \ \delta_{ss'}(1-\lambda_{st}). \label{eq:appendy4}
\end{eqnarray}
Note that, from (\ref{eq:appen1}), $r_{0st} = 1$ and, from (\ref{eq:appen2}), $r_{st} = \lambda_{st}(1-\lambda_{st})^{-1}$ so that $r_{Nst} = r_{0st} + r_{st}$ follows immediately from (\ref{eq:appendy4}). Similarly, using (\ref{eq:adj1}),
\begin{eqnarray}
E_{\overline{\mathcal{W}}_{t}}(\mathcal{M}(Y_{st})) & = & E(\mathcal{M}(Y_{st})) + V_{st}^{T}A\{A + \phi_{t}^{-1}N^{-1}B - N^{-1}\hat{A}\}^{-1}\{\overline{\mathcal{W}}_{t} - E(\overline{\mathcal{W}}_{t})\} \nonumber \\
& = & E(\mathcal{M}(Y_{st})) + \lambda_{st}V_{st}^{T}\{\overline{\mathcal{W}}_{t} - E(\overline{\mathcal{W}}_{t})\} \nonumber \\
& = & E(\mathcal{M}(Y_{st})) + \lambda_{st}\{\overline{Y}_{st} - E(\mathcal{M}(Y_{st}))\} \label{eq:appendy5}
\end{eqnarray}
as $E(\overline{\mathcal{C}}(N)) = E(\mathcal{M}(\mathcal{C}))$. $\mu_{Nst}$ follows by suitable rearrangement of (\ref{eq:appendy5}). \hfill $\Box$

\ \newline
\textbf{Proof of Lemma \ref{lem:sep} -} Noting that $\mathcal{M}(\mathcal{W}_{t}) = (I_{g_{0}} \otimes U_{t}^{T})\mathcal{M}(\mathcal{C})$ then, from Assumption \ref{assum2} and Definition \ref{def:var}, for all $t$, $t'$ we have 
$Cov(\mathcal{M}(\mathcal{W}_{t}),\mathcal{M}(\mathcal{W}_{t'})) = \delta_{tt'}A$ and $Cov(\mathcal{M}(\mathcal{W}_{t}), \mathcal{M}(\mathcal{C})) = A \otimes U_{t}^{T}C$. Thus, $Var(\mu(\mathcal{W})) = \oplus_{t=1}^{v_{0}} A$ and, noting that $\mathcal{M}(W_{gt}) \in \langle \mathcal{M}(\mathcal{C}_{g}) \rangle$ so that $Cov(\overline{\mathcal{C}}(N), \mathcal{M}(\mathcal{W}_{t})) = Cov(\mathcal{M}(\mathcal{C}), \mathcal{M}(\mathcal{W}_{t}))$, $Cov(\overline{\mathcal{C}}(N), \mathcal{M}(\mathcal{W})) = Cov(\mathcal{M}(\mathcal{C}), \mathcal{M}(\mathcal{W})) = [A \otimes CU_{1} \, \ldots \, A \otimes CU_{v_{0}}]$. Hence,
\begin{eqnarray}
Cov(\overline{\mathcal{C}}(N), \mathcal{M}(\mathcal{W}))Var^{-1}(\mathcal{M}(\mathcal{W}))Cov(\mathcal{M}(\mathcal{W}), \mathcal{M}(\mathcal{C})) & = & A \otimes \sum_{v=1}^{v_{0}} CU_{v}U_{v}^{T}C \nonumber \\
  = \ A \otimes CUU^{T}C & = & Cov(\overline{\mathcal{C}}(N), \mathcal{M}(\mathcal{C})) \label{eq:append1}
\end{eqnarray}
where (\ref{eq:append1}) follows as $U^{T}CU = I$ implies that $UU^{T}C = I$. Let $\tilde{D} = [I_{g_{0}} \otimes U_{1} \, \ldots \, I_{g_{0}} \otimes U_{v_{0}}]$. Then, by direct multiplication and Definition \ref{def:var},
\begin{eqnarray*}
 \tilde{D}^{T}Var(\overline{\mathcal{C}}(N))\tilde{D} & = & \oplus_{t=1}^{v_{0}} \{A + \phi_{t}^{-1} N^{-1}B - N^{-1}\hat{A}\}.
\end{eqnarray*}
Inverting and rearranging gives
\begin{eqnarray}
Var^{-1}(\overline{\mathcal{C}}(N)) & = & \tilde{D}\left[\oplus_{t=1}^{v_{0}} \{A + \phi_{t}^{-1} N^{-1}B - N^{-1}\hat{A}\}^{-1}\right]\tilde{D}^{T}. \label{eq:appendix1}
\end{eqnarray}
Now, $Cov(\mathcal{M}(\mathcal{W}_{t}), \overline{\mathcal{C}}(N))\tilde{D} = (A \otimes U_{t}^{T}C)\tilde{D} = [\delta_{t1}A \, \ldots \, \delta_{tv_{0}}A]$ so that, using (\ref{eq:appendix1}),
\begin{eqnarray}
\lefteqn{Cov(\mathcal{M}(\mathcal{W}_{t}), \overline{\mathcal{C}}(N))Var^{-1}(\overline{\mathcal{C}}(N)) \ =} \nonumber \\ & & \ \ \left[\delta_{t1}A \{A + \phi_{1}^{-1} N^{-1}B - N^{-1}\hat{A}\}^{-1}\, \ldots \, \delta_{tv_{0}}A\{A + \phi_{v_{0}}^{-1} N^{-1}B - N^{-1}\hat{A}\}^{-1}\right]\tilde{D}^{T} \label{eq:appendix2}
\end{eqnarray}
and 
\begin{eqnarray}
Cov(\mathcal{M}(\mathcal{W}_{t}), \overline{\mathcal{C}}(N))Var^{-1}(\overline{\mathcal{C}}(N))Cov(\overline{\mathcal{C}}(N), \mathcal{M}(\mathcal{W}_{t'})) \ =  \hspace{3.5cm}\nonumber \\ \delta_{tt'} A\{A + \phi_{t}^{-1}N^{-1}B - N^{-1}\hat{A}\}^{-1}A. \label{eq:appendix3}
\end{eqnarray}
From (\ref{eq:appen1}) and (\ref{eq:appendix3}) it follows that for all $t \neq t'$,
\begin{eqnarray}
Cov(\mathcal{M}(\mathcal{W}_{t}), \mathcal{M}(\mathcal{W}_{t'})) \ = \ Cov(\mathcal{M}(\mathcal{W}_{t}), \overline{\mathcal{C}}(N))Var^{-1}(\overline{\mathcal{C}}(N))Cov(\overline{\mathcal{C}}(N), \mathcal{M}(\mathcal{W}_{t'})). \label{eq:appen6}
\end{eqnarray}
Results 1. and 2. follow, respectively, from (\ref{eq:append1}) and (\ref{eq:appen6}) using Theorems 5.20 and 5.23 of Goldstein and Wooff (2007).\nocite{Goldstein/Wooff:2007} \hfill $\Box$ 

\ \newline
\textbf{Proof of Theorem \ref{theo:groupfin} -} Noting that $Cov(\widetilde{\mathcal{M}}(\mathcal{C}), \overline{\mathcal{C}}(N)) = Var(\widetilde{\mathcal{M}}(\mathcal{C})) = (A \otimes C) + (M^{-1}B \otimes D) - (M^{-1}\hat{A} \otimes C)$ then, as $Var(\overline{\mathcal{C}}(N)) =  (A \otimes C) + (N^{-1}B \otimes D) - (N^{-1}\hat{A} \otimes C)$, we have $Var_{\widetilde{\mathcal{M}}(\mathcal{C})}(\overline{\mathcal{C}}(N)) = \{(N^{-1} - M^{-1})B \otimes D\} - \{(N^{-1} - M^{-1})\hat{A} \otimes C)\}$. Now, $\widetilde{\mathcal{M}}(Z_{st}) = \tilde{V}_{st}^{T}\widetilde{\mathcal{M}}(\mathcal{W}_{t}) = (\tilde{V}_{st}^{T} \otimes U_{t}^{T})\widetilde{\mathcal{M}}(\mathcal{C})$ and $\overline{Z}_{st} = \tilde{V}_{st}^{T}\overline{\mathcal{W}}_{t} = (\tilde{V}_{st}^{T} \otimes U_{t}^{T})\overline{\mathcal{C}}(N)$, so that from Definitions \ref{def:var} and \ref{fincangroup}, for all $s, s', t, t'$ we have 
\begin{eqnarray}
Cov(\widetilde{\mathcal{M}}(Z_{st}), \widetilde{\mathcal{M}}(Z_{s't'})) & = & (\tilde{V}_{st}^{T} \otimes U_{t}^{T})Var(\widetilde{\mathcal{M}}(\mathcal{C}))(\tilde{V}_{s't'} \otimes U_{t'}) \nonumber \\
& = & \delta_{tt'}\tilde{V}_{st}^{T}\{A + \phi_{t}^{-1}M^{-1}B - M^{-1}\hat{A}\}\tilde{V}_{s't} \ = \ \delta_{ss'}\delta_{tt'}, \label{eq:appen1fin} \\
Cov_{\widetilde{\mathcal{M}}(\mathcal{C})}(\overline{Z}_{st}, \overline{Z}_{s't'}) & = & \delta_{tt'}\tilde{V}_{st}^{T}\{\phi_{t}^{-1}(N^{-1} - M^{-1})B - (N^{-1} - M^{-1})\hat{A}\}\tilde{V}_{s't} \nonumber \\
& = & \delta_{ss'}\delta_{tt'} (1 - \tilde{\lambda}_{st})\tilde{\lambda}_{st}^{-1}. \label{eq:appen2fin}
\end{eqnarray}
Now as $Cov(\widetilde{\mathcal{M}}(\mathcal{W}_{t}), \overline{\mathcal{W}}_{t}) = Var(\widetilde{\mathcal{M}}(\mathcal{W}_{t})) = A + \phi_{t}^{-1}M^{-1}B - M^{-1}\hat{A}$ and $Var(\overline{\mathcal{W}}_{t})) = A + \phi_{t}^{-1}N^{-1}B - N^{-1}\hat{A}$ then, from (\ref{eq:adj3}), the resolution transform for the adjustment of $\langle \widetilde{\mathcal{M}}(\mathcal{W}_{t}) \rangle$ given $\overline{\mathcal{W}}_{t}$ is,
\begin{eqnarray}
T_{\widetilde{\mathcal{M}}(\mathcal{W}_{t}):\overline{\mathcal{W}}_{t}} & = & \{A + \phi_{t}^{-1}N^{-1}B- N^{-1}\hat{A}\}^{-1}\{A + \phi_{t}^{-1}M^{-1}B- M^{-1}\hat{A}\} \label{eq:appendix4fin}
\end{eqnarray}
which, from Definition \ref{fincangroup}, for each $s = 1, \ldots, g_{0}$, has eigenvector $\tilde{V}_{st}$ and corresponding eigenvalue $\tilde{\lambda}_{st}$. In a similar way to (\ref{eq:appendy4}) and (\ref{eq:appendy5}) we may deduce that
\begin{eqnarray}
Cov_{\overline{\mathcal{W}}_{t}}(\widetilde{\mathcal{M}}(Z_{st}), \widetilde{\mathcal{M}}(Z_{s't})) & = & \delta_{ss'}(1-\tilde{\lambda}_{st}), \label{fin:appendy4} \\
E_{\overline{\mathcal{W}}_{t}}(\widetilde{\mathcal{M}}(Z_{st})) & = & E(\widetilde{\mathcal{M}}(Z_{st})) + \tilde{\lambda}_{st}\{\overline{Z}_{st} - E(\widetilde{\mathcal{M}}(Z_{st}))\} \label{fin:appendy5}
\end{eqnarray}
so that $\tilde{r}_{Nst}$ and $\tilde{\mu}_{Nst}$ follow, respectively, from (\ref{fin:appendy4}) and (\ref{fin:appendy5}) using (\ref{eq:appen1fin}) and (\ref{eq:appen2fin}) when $s = s'$ and $t = t'$. \hfill $\Box$

\ \newline
\textbf{Proof of Lemma \ref{lem:sepfin} -} By writing $\widetilde{\mathcal{M}}(W_{gt}) = U_{t}^{T} \widetilde{\mathcal{M}}(\mathcal{C}_{g})$ then for all $g$, $h$, $t$, $t'$ we have $Cov(\widetilde{\mathcal{M}}(W_{gt}), \widetilde{\mathcal{M}}(W_{ht'})) = \{\alpha_{gh} +\delta_{gh}\frac{1}{m_{g}}(\phi_{t}^{-1}\gamma_{g} - \alpha_{gg})\}\delta_{tt'}$. As a result, $Var(\widetilde{\mathcal{M}}(\mathcal{W})) = \oplus_{t=1}^{v_{0}} \{A + \phi_{t}^{-1}M^{-1}B - M^{-1}\hat{A}\}$. Now $Cov(\widetilde{\mathcal{M}}(\mathcal{C}), \widetilde{\mathcal{M}}(\mathcal{W})) = [(A \otimes CU_{1}) + (M^{-1}B \otimes DU_{1}) - (M^{-1}\hat{A} \otimes CU_{1})\, \ldots \, (A \otimes CU_{v_{0}}) + (M^{-1}B \otimes DU_{v_{0}}) - (M^{-1}\hat{A} \otimes CU_{v_{0}})] = [\{A + \phi_{1}^{-1}M^{-1}B - M^{-1}\hat{A}\} \otimes CU_{1} \, \ldots \, \{A + \phi_{v_{0}}^{-1}M^{-1}B - M^{-1}\hat{A}\} \otimes CU_{v_{0}}]$. Hence, by direct multiplication using that $Cov(\overline{\mathcal{C}}(N), \widetilde{\mathcal{M}}(\mathcal{W})) = Cov(\widetilde{\mathcal{M}}(\mathcal{C}), \widetilde{\mathcal{M}}(\mathcal{W}))$,
\begin{eqnarray}
Cov(\overline{\mathcal{C}}(N), \widetilde{\mathcal{M}}(\mathcal{W}))Var^{-1}(\widetilde{\mathcal{M}}(\mathcal{W}))Cov(\widetilde{\mathcal{M}}(\mathcal{W}), \widetilde{\mathcal{M}}(\mathcal{C})) & = & (A \otimes CUU^{T}C) +  \nonumber \\
 (M^{-1}B \otimes DUU^{T}C) - (M^{-1}\hat{A} \otimes CUU^{T}C) & = & Cov(\overline{\mathcal{C}}(N), \widetilde{\mathcal{M}}(\mathcal{C})) \label{fin:append1}
\end{eqnarray}
where (\ref{fin:append1}) follows in a identical way to (\ref{eq:append1}). Now, as $Cov(\widetilde{\mathcal{M}}(\mathcal{W}_{t}), \overline{\mathcal{C}}(N)) = \{A + \phi_{t}^{-1}M^{-1}B-M^{-1}\hat{A}\} \otimes U^{T}_{t}C$, then using an equivalent method to (\ref{eq:appendix3}) we have
\begin{eqnarray}
Cov(\widetilde{\mathcal{M}}(\mathcal{W}_{t}), \overline{\mathcal{C}}(N))Var^{-1}(\overline{\mathcal{C}}(N))Cov(\overline{\mathcal{C}}(N), \widetilde{\mathcal{M}}(\mathcal{W}_{t'})) \ =  \hspace{4.5cm}\nonumber \\ \delta_{tt'} \{A + \phi_{t}^{-1}M^{-1}B-M^{-1}\hat{A}\}\{A + \phi_{t}^{-1} N^{-1}B-N^{-1}\hat{A}\}^{-1}\{A + \phi_{t}^{-1}M^{-1}B-M^{-1}\hat{A}\}. \label{fin:appendix3}
\end{eqnarray}
From (\ref{eq:appen1fin}) and (\ref{fin:appendix3}) it follows that for all $t \neq t'$,
\begin{eqnarray}
Cov(\widetilde{\mathcal{M}}(\mathcal{W}_{t}), \widetilde{\mathcal{M}}(\mathcal{W}_{t'})) \ = \ Cov(\widetilde{\mathcal{M}}(\mathcal{W}_{t}), \overline{\mathcal{C}}(N))Var^{-1}(\overline{\mathcal{C}}(N))Cov(\overline{\mathcal{C}}(N), \widetilde{\mathcal{M}}(\mathcal{W}_{t'})). \label{fin:appen6}
\end{eqnarray}
Results 1. and 2. follow, respectively, from (\ref{fin:append1}) and (\ref{fin:appen6}) using Theorems 5.20 and 5.23 of Goldstein and Wooff (2007).\nocite{Goldstein/Wooff:2007} \hfill $\Box$ 

\end{document}